\documentclass[11pt,a4paper]{article}
\usepackage[T1]{fontenc}
\usepackage[utf8]{inputenc}
\usepackage{jheppub}
\usepackage{amsfonts,mathtools,braket,slashed,bm,bbm}
\usepackage[table,dvipsnames]{xcolor}
\definecolor{Lightgray}{gray}{0.93}
\usepackage[font=small,labelfont=bf,labelsep=period]{caption}
\usepackage{subcaption}
\usepackage{cancel}
\usepackage{booktabs}
\usepackage[section]{placeins}
\usepackage[compat=1.1.0]{tikz-feynman}
\usetikzlibrary{arrows.meta,decorations.markings}
\hypersetup{bookmarksnumbered=true,bookmarksopen=true}
\usepackage{marginnote}
\usepackage[normalem]{ulem}
\renewcommand{\afterTocRuleSpace}{\clearpage}
\allowdisplaybreaks
\def\ff{f\hspace{-0.15cm}f}

\newcommand{\as}{\alpha_s}

\newcommand{\df}{\mathrm{d}}

\newcommand{\cH}{\mathcal{H}}

\newcommand{\cO}{\mathcal{O}}
\newcommand{\cS}{\mathcal{S}}

\newcommand{\Horig}{\bm{\mathcal H}}
\newcommand{\Hh}{\widehat{\bm{\mathcal H}}}

\title{Resummation of Next-to-Leading Non-Global Logarithms in Higgs Production}
\author[a]{Thomas Becher,}
\author[a]{Rebecca von Kuk,}
\author[a,b,c]{Xinyu Qin,}
\author[d]{and Nicolas Schalch}

\affiliation[a]{Albert Einstein Center for Fundamental Physics, Institut f\"ur Theoretische Physik,\\
Universit\"at Bern, Sidlerstrasse 5, CH-3012 Bern, Switzerland}

\affiliation[b]{Department of Physics, P.O. Box 35, FI-40014 University of Jyväskylä, Finland}

\affiliation[c]{Helsinki Institute of Physics, P.O. Box 64, FI-00014 University of Helsinki, Finland}

\affiliation[d]{Rudolf Peierls Centre for Theoretical Physics, Clarendon Laboratory,\\
Parks Road, OX1 3PU, Oxford, UK}

\emailAdd{becher@itp.unibe.ch}
\emailAdd{rebecca.vonkuk@unibe.ch}
\emailAdd{qinx@jyu.fi}
\emailAdd{nicolas.schalch@physics.ox.ac.uk}
\abstract{We analyze Higgs production at the LHC with a jet veto at central rapidities and present the first next-to-leading logarithmic resummation of non-global logarithms for this process. The resummation combines one-loop hard and soft
matching functions with two-loop anomalous dimensions in the large-$N_c$ limit. To perform the resummation we numerically solve 
renormalization group equations via the parton-shower framework {\sc Marzili}. After matching to fixed order, we perform a detailed study of different sources of uncertainties. We find that the hard matching corrections dominate by far and are substantially larger than in Drell--Yan production. Together with earlier results for Drell--Yan, these are the first resummations of subleading non-global logarithms in hadron-collider cross sections, and they provide a benchmark for logarithmically accurate parton showers.}

\begin{document}
\maketitle

\section{Introduction}
Vetoes on hadronic radiation are important in precision studies of Higgs-boson production at the LHC, where they define exclusive event categories and suppress backgrounds. A classic example is the fully leptonic $H\to WW^*\to \ell\nu\ell\nu$ channel, whose dominant background is top-pair production: each top quark decays as $t\to Wb$, so even when both $W$ bosons decay leptonically the event typically contains two $b$-initiated jets. To deal with the top-pair background, experimental measurements split their events into jet bins, and the zero-jet bin has the highest signal purity \cite{ATLAS:2022ooq,CMS:2022uhn,ATLAS:2025abg,ATLAS:2025hki,CMS:2026igg}.
A veto on additional jets induces a low scale $Q_0$, and fixed-order predictions contain terms enhanced by powers of the logarithm $L=\ln(m_H/Q_0)$. For $Q_0\ll m_H$, a reliable description requires their all-order resummation. The resummation for Higgs production in the zero-jet bin, i.e.\ the cross section with a jet veto, has reached high perturbative accuracy~\cite{Banfi:2012yh,Banfi:2012jm,Becher:2012qa,Becher:2013xia,Becher:2014aya,Banfi:2015pju,Stewart:2013faa,Dawson:2016ysj,Campbell:2023cha,Gavardi:2023aco}. These resummations assume that the jet veto is imposed globally, at all rapidities, while experimental measurements have finite rapidity coverage. A non-global veto on radiation leads to additional logarithmic enhancements, known as non-global logarithms (NGLs)~\cite{Dasgupta:2001sh}. The size of these effects was estimated to be small in the zero-jet bin since the rapidity coverage is large enough to suppress them~\cite{Banfi:2012yh,Balsiger:2018ezi}. The situation is different in the one-jet bin: the presence of a jet at central rapidities implies that the observable is genuinely non-global and the resulting NGLs are not fully accounted for in existing resummations~\cite{Liu:2012sz,Liu:2013hba,Cal:2024yjz}.

In this paper we study NGLs in Higgs production in a simple setting: we impose a veto on the total transverse energy deposited in a finite rapidity slice centered on the Higgs boson. Specifically, we study the corresponding gap fraction,
\begin{align}
  R(Q_0)=\frac{\sigma(E_T^{\rm gap}<Q_0)}{\sigma_{\rm incl}}\,,
  \label{eq:gap_fraction}
\end{align}
in Higgs production via gluon fusion. At NLO, a constraint on the energy $E_T^{\rm gap}$ coincides with a jet-$p_T$ veto in the same rapidity interval; beyond this order, soft clustering effects must be taken into account.
Because the radiation constraint is imposed in a limited angular region of phase space, the gap fraction is a classic example of a non-global observable~\cite{Dasgupta:2002bw}. An emission outside the gap acts as a new color source for softer radiation into the gap, and the resulting real--virtual mismatch correlates emissions at different angles and energy scales. A simple Sudakov factor, based on a fixed set of radiators, cannot capture this branching structure: as radiation is generated, the number, directions and colors of the emitters evolve. The corresponding evolution couples configurations of arbitrarily high parton multiplicity and generally requires a numerical or semi-numerical treatment, typically based on Monte Carlo methods. The non-global logarithms considered here arise from soft wide-angle radiation. For a fixed rapidity interval, that is excluding the beams, the soft evolution has a leading-logarithmic (LL) tower of single logarithms $\alpha_s^nL^n$ in the planar limit. This structure was identified in~\cite{Dasgupta:2002bw}, formulated as a large-$N_c$ evolution equation in~\cite{Banfi:2002hw}, and later recast as renormalization-group (RG) evolution in effective field theory~\cite{Becher:2015hka,Becher:2016mmh}.

Only in the past few years has the resummation of the next-to-leading logarithmic (NLL) series, $\alpha_s^n L^{n-1}$, become possible. Complementary formulations based on generating functionals~\cite{Banfi:2021owj,Banfi:2021xzn}, logarithmically accurate parton showers~\cite{Dasgupta:2020fwr,FerrarioRavasio:2023kyg}, and renormalization-group evolution in effective field theory (EFT)~\cite{Becher:2021urs,Becher:2023vrh} have provided independent realizations of higher-order soft evolution. 
Agreement among these approaches for the transverse energy flow within a rapidity slice in $e^+e^-$ annihilation provides a non-trivial validation of the NLL resummation~\cite{GnoleComparison}. Among these frameworks, the EFT approach was the first to be extended, at subleading accuracy, to a non-global observable at a hadron collider~\cite{Becher:2023vrh}. This demonstrates the strength of separating universal soft evolution from process-dependent hard scattering: the same evolution can be applied across different collider environments, with the hard matching functions supplying the process-specific information. In this work, we exploit this structure to extend NLL non-global resummation to Higgs production via gluon fusion. The leading-power hadronic cross section takes the schematic form~\cite{Becher:2021zkk,Becher:2023mtx,Becher:2025igg}
\begin{align}
\sigma(Q_0)=\sum_{ij}\int\!dx_1dx_2\sum_{m=0}^{\infty}
 \big\langle \Horig_{ij\to m}(m_H,\mu)\otimes
 \bm{\mathcal W}_{ij\to m}(Q_0,\mu) \big\rangle\,.
 \label{eq:intro_factorization}
\end{align}
The partonic hard functions $\Horig_{ij\to m}$ describe the short-distance scattering of partons $i$ and $j$ into an $m$-parton final state, and $x_{1,2}$ denote the longitudinal momentum fractions of the partons entering the hard process. The low-energy matrix elements $\bm{\mathcal W}_{ij\to m}$ contain soft Wilson lines along the directions of the hard partons and collinear fields for the incoming partons. The symbol $\otimes$ indicates the integral over the directions of the final-state hard partons which is performed after the combination with the low-energy matrix elements $\bm{\mathcal W}_{ij\to m}$; the brackets $\langle \dots \rangle$ denote a color trace. The soft and collinear partons in the low-energy matrix elements $\bm{\mathcal W}_{ij\to m}$ can interact via the exchange of Glauber gluons. These Glauber interactions vanish in the large-$N_c$ limit adopted in this paper, and the low-energy matrix elements then factorize into parton distribution functions (PDFs) and soft Wilson-line matrix elements, as will be detailed in Section~\ref{sec:hard_functions} below.

RG evolution of the hard functions from $\mu_h\sim m_H$ to $\mu_s\sim Q_0$ resums the non-global logarithms; the evolution increases parton multiplicities and we solve it numerically with the large-$N_c$ parton-shower framework \textsc{Marzili}~\cite{Becher:2023vrh}. At NLL accuracy, the two-loop anomalous dimension, capturing the higher-order soft behavior, must be combined with one-loop hard and soft matching corrections. Universal higher-order soft evolution is therefore essential but not sufficient: hard matching supplies the exact resolved-radiation pattern, soft matching the finite measurement-dependent terms, and fixed-order matching the power corrections away from the soft limit.

\looseness=-1 The first NLL application of this framework at a hadron collider concerned the analogous gap fraction in Drell--Yan production~\cite{Becher:2023vrh}. Extending it to gluon-fusion Higgs production is non-trivial because the Born process contains two incoming gluons and the NLO real corrections involve the $gg$, $qg$, $gq$, and $q\bar q$ channels. However, in the planar limit, an adjoint gluon line decomposes into two fundamental color lines. The required soft evolution can then be assembled from products or direct analogues of the Drell--Yan building blocks. The only genuinely new inputs for Higgs production are the NLO hard matching corrections.

The hard functions required for \eqref{eq:intro_factorization} are directly related to NLO partonic cross sections. However, they cannot be fully reconstructed from the existing literature~\cite{Anastasiou:2002qz}, since fixed-order computations generally assume infrared safe observables to simplify the partonic results, e.g.\ by inclusively integrating purely soft terms over phase space. The factorization formula \eqref{eq:intro_factorization} separates hard and soft emissions, and the hard emissions are not infrared safe by themselves, so these simplifications are not appropriate.
We derive the one-loop two-parton and tree-level three-parton hard functions in the heavy-top effective theory and use them to match the resummed result to the exact NLO prediction.

The main result of our paper is the first large-$N_c$ NLL resummation of non-global logarithms for the central-gap fraction in Higgs production, combined with the full NLO prediction in the heavy-top effective theory. As for the inclusive Higgs cross section, the hard functions in \eqref{eq:intro_factorization} receive very large NLO corrections, so that the NLL corrections to the cross section are substantial. Part of these corrections cancel in the ratio \eqref{eq:gap_fraction}, but the cancellation sensitively depends on the expansion scheme adopted in the computation of the gap fraction, as in the case of the inclusive jet-veto efficiency \cite{Banfi:2012yh}.

Our calculation provides a benchmark for logarithmically accurate parton showers and a step towards non-global NLL resummation for exclusive Higgs-plus-jet observables, where radiation inside the resolved jet is allowed while radiation outside it is vetoed. This extension requires additional hard matching functions and an explicit treatment of jet clustering effects. The extension of the factorization theorem \eqref{eq:intro_factorization} to account for clustering jets, as well as the relevant two-loop anomalous dimensions are now available~\cite{Becher:2023znt,Becher:2026zon}.

The remainder of the paper is organized as follows. In Section~\ref{sec:hard_functions} we derive the hard matching functions. Section~\ref{sec:resummation} describes the NLL evolution and its relation to the Drell--Yan soft functions. In Section~\ref{sec:numerical_results} we present the matched NLL+NLO predictions, together with a detailed analysis of different sources of uncertainties, before concluding in Section~\ref{sec:conclusions}. In a series of appendices, we present benchmark numbers and perform additional scale-variation studies.

\section{Hard functions}
\label{sec:hard_functions}

We first specify the hadronic factorization and the conventions for the hard functions, including their PDF convolution. We then derive the NLO Higgs-production kernels, retaining the angular dependence required by the gap measurement.

\subsection{Hadronic factorization and conventions}
The factorization theorem~\eqref{eq:intro_factorization} separates the hard interaction from the low-energy dynamics of the two incoming hadrons. We write their momenta as $P_1$ and $P_2$, with $s=(P_1+P_2)^2$, and the incoming parton momenta as $p_1=x_1P_1$ and $p_2=x_2P_2$. The indices $i,j$ label their flavors. The multiplicity $m$ counts the outgoing colored hard partons; in addition there are the two incoming partons. The color-singlet Higgs boson is not counted. The hard process thus has $m=0$ at Born level and $m=1$ for the real-emission channels at NLO.

The angular region in which we do not restrict radiation is defined by two cones around the beams, in the spirit of the fixed-cone definition of Sterman and Weinberg~\cite{Sterman:1977wj}.  The constraint $\Theta_{\rm in}$ requires energetic final-state partons to lie in one of the cones, which are chosen symmetric in the frame where the Higgs boson has vanishing rapidity. The veto region is the complement, i.e.\ the central slice $|Y|<Y_{\max}$, with $\Delta Y=2Y_{\max}$. These are fixed angular regions, not jets reconstructed with a sequential recombination algorithm. At NLO, the energy-flow measurement of~\eqref{eq:gap_fraction} is equivalent to a measurement of the energy of the leading jet in the veto region. Beyond this order, the soft functions will differ.

The matrix element $\bm{\mathcal W}_{ij\to m}$ contains incoming-collinear fields as well as soft Wilson lines along all $(m+2)$ colored directions. However, in the large-$N_c$ limit used in this work, the incoming-collinear dynamics factorizes into ordinary unpolarized PDFs, leaving
\begin{align}
 \bm{\mathcal W}_{ij\to m}
 &\longrightarrow f_{i/P_1}(x_1,\mu_f)\,
 f_{j/P_2}(x_2,\mu_f)\,
 \bm{\mathcal S}_{ij\to m}(\{\underline{n}\},Q_0,\mu)\,.
 \label{eq:W_to_PDFs}
\end{align}
The soft function $\bm{\mathcal S}_{ij\to m}$ is the vacuum matrix element of the incoming and outgoing Wilson lines with constraint $E_T^{\rm gap}<Q_0$. It depends on their color representations and directions, as well as on the geometry of the veto region, but not on the Higgs-production amplitude. The large-$N_c$ factorization \eqref{eq:W_to_PDFs} also implies that the collinear and soft matrix elements evolve independently of each other, and we keep the factorization scale $\mu_f$ in the PDF distinct from the renormalization scale in the soft function. 

At finite $N_c$, the evolution of the low-energy matrix elements is significantly more complicated: both the soft and collinear evolution have nontrivial color structure and do not commute due to the presence of Glauber phases. The soft-collinear factorization \eqref{eq:W_to_PDFs} breaks down and the evolution generates super-leading logarithms (SLLs), whose resummation has been studied in~\cite{Becher:2021zkk,Becher:2023mtx,Becher:2024nqc}. Perturbative Glauber contributions to the low-energy matrix elements reconcile this evolution with PDF factorization below $Q_0$~\cite{Becher:2024kmk,Becher:2025igg}. Coherence-violating effects also occur for global observables~\cite{Becher:2026kbr}. These effects are not included in the (planar) NLL predictions presented here.
While SLLs can be sizable in processes with final-state jets \cite{Becher:2024nqc}, their numerical effect appears to be small for Higgs-boson production \cite{Becher:2023mtx}. They start one order higher than for jet processes since, compared to that case, an additional emission is needed to generate the necessary color structure. For veto scales around $20\,\mathrm{GeV}$ and a gap size of $\Delta Y=2$ the leading SLLs were found to be below $0.5 \%$ in \cite{Becher:2023mtx}.

At fixed incoming momenta, the bare partonic hard function is the color density matrix for $ij\to H+m$ partons, integrated over their energies and the Higgs momentum while retaining the colored directions:
\begin{align}
 \Horig_{ij\to m}(\{\underline{n}\},p_1,p_2)
 ={}&\frac{1}{2\hat s}\int\frac{d^{d-1}q_H}{(2\pi)^{d-1}\,2E_H}
 \prod_{k=3}^{m+2}\int\frac{dE_k\,E_k^{d-3}}{\tilde c^{\epsilon}(2\pi)^2}
 \nonumber\\
 &\times|\mathcal M_{ij\to m}\rangle\langle\mathcal M_{ij\to m}|
 \,(2\pi)^d\,\delta^{(d)}\!\left(p_1+p_2-q_H-\sum_{k=3}^{m+2}E_kn_k\right)
 \Theta_{\rm in}.
 \label{eq:hard_bare_hadronic}
\end{align}
Here $\hat s=(p_1+p_2)^2$, $q_H^2=m_H^2$, $E_H=q_H^0$, and $p_k=E_kn_k$. The delta function imposes total energy--momentum conservation, including the Higgs momentum. The normalization of the energy integrals follows Ref.~\cite{Becher:2023vrh}, with $d=4-2\epsilon$ and $\tilde c=e^{\gamma_E}/\pi$; the directions $n_k$ are integrated only in the convolution $\otimes$. Spin sums, initial-state averages and final-state symmetry factors are understood. At higher multiplicities, there are different $m$-parton states, which must be distinguished and have separate hard functions. 

After PDF factorization, the cross section can be written in the usual way, as a convolution of PDFs with a partonic cross section
\begin{align}
 \sigma(Q_0)
 =\sum_{ij}\int dx_1dx_2\,
 f_{i/P_1}(x_1,\mu_f) f_{j/P_2}(x_2,\mu_f)\,
 \hat{\sigma}_{ij}(z,Q_0)\,, 
 \label{eq:PDF_hard}
\end{align}
where
\begin{align}
z&=\frac{m_H^2}{\hat s}=\frac{\tau}{x_1x_2}  ,&\text{ with }& & \tau& =\frac{m_H^2}{s} \label{eq:zdef}\,.
\end{align}
We define the parton luminosity as
\begin{align}
 \ff_{ij}(\xi,\mu_f)=\int_{\xi}^{1}\frac{dx}{x}\,
 f_{i/P_1}(x,\mu_f)f_{j/P_2}(\xi/x,\mu_f)\,
 \label{eq:luminosity}
\end{align}
and rewrite the hadronic cross section as a convolution of the partonic cross section with the luminosity,
\begin{align}
\sigma(Q_0) =\sum_{ij}\int_{\tau}^{1}dz\,\frac{\tau}{z^2}\,
 \ff_{ij}(\tau/z,\mu_f)\hat{\sigma}_{ij}(z,Q_0) \,.
 \label{eq:luminosity_convolution}
\end{align}

\subsection{Partonic hard functions at NLO}

We now compute the NLO partonic hard functions for Higgs boson production via gluon fusion in the presence of an angular veto. We retain the top-quark contribution and neglect the lighter-quark loops.
In the heavy top-quark limit ($m_t \to \infty$), the top quark can be integrated out and the dynamics are described by the Lagrangian of Higgs effective theory (HEFT)~\cite{Wilczek:1977zn,Shifman:1979eb, Inami:1982xt} 
\begin{align}
    \mathcal{L}_{\text{HEFT}} = \frac{\alpha_s}{4\pi}  \frac{C_t}{ 3v}  H G_{\mu\nu}^a G^{a,\mu\nu} \,,
\end{align}
where $v$ is the vacuum expectation value, $G_{\mu\nu}^a$ is the gluon field strength tensor, and $C_t$ is the Wilson coefficient arising when integrating out the top quark.
This Wilson coefficient is known up to high orders in QCD \cite{Djouadi:1991tka,DAWSON1991283,Chetyrkin:1997iv,  vanRitbergen:1997va, Czakon:2004bu, Chetyrkin:2005ia, Schroder:2005hy, Gerlach:2018hen}. 
However, for our purposes it is sufficient to consider the NLO result which in the $\overline{\text{MS}}$ scheme reads
\begin{align}
    C_t = 1 + \frac{\alpha_s}{4\pi}(5C_A-3C_F) + \mathcal{O}(\alpha_s^2)\,. \label{eq:Ct_NLO}
\end{align}
At leading order, the only relevant partonic process is given by $gg \to H$ and the corresponding cross section in HEFT is 
\begin{align}
    \hat{\sigma}^{\mathrm{LO}}_{gg} \equiv \sigma_0 \,\delta(1-z),
\end{align}
with the variable $z$ defined in~\eqref{eq:zdef}. The $d$-dimensional Born normalization is
\begin{align}
    \sigma_0 = \frac{\alpha_s^2}{576\pi v^2} \frac{1}{1-\epsilon}\, .\label{eq:Born_norm_d}
\end{align}
In finite, renormalized expressions we can set $\epsilon=0$ in $\sigma_0$.

To separate matching at the top-quark scale from the subsequent QCD dynamics, we factor out the common Wilson coefficient as well as the tree-level normalization and define the reduced hard functions
\begin{align}
 \Horig_{ij\to m}
 =\sigma_0\,|C_t|^2\,\Hh_{ij\to m}
 \label{eq:Ct_factorization}\, .
\end{align}
In terms of these, the partonic cross section reads
\begin{align}
 \hat{\sigma}_{ij}(z,Q_0)
 = \sigma_0\, |C_t|^2\, \sum_{m\geq 0}
 \langle \Hh_{ij\to m} \otimes\bm{\mathcal S}_{ij\to m}\rangle.
 \label{eq:reduced_factorized_cross_section}
\end{align}

With this normalization, the hard and soft functions required for the NLL+NLO prediction are expanded through first order relative to the Born cross section:
\begin{align}
   \Hh_{gg\to 0} &= \bm{\Hh}_{2,gg}^{(0)} + \frac{\as}{4 \pi} \bm{\Hh}_{2,gg}^{(1)} + \dots \,, \\
  \Hh_{ij\to 1} &=  \frac{\as}{4 \pi} \bm{\Hh}_{3,ij}^{(1)} + \dots , \\
    \bm{\cS}_{ij\to m} &= \mathbf{1}+ \frac{\as}{4 \pi} \bm{\cS}_{(m+2),ij}^{(1)} + \dots \,,
    \label{eq:hard_soft_expansions}
\end{align}
where the subscript on the expansion coefficients indicates the total number of hard partons and the superscript the order in $\alpha_s$ relative to the Born cross section. 

The NLO fixed-order expansion of the partonic cross section \eqref{eq:reduced_factorized_cross_section} in the $gg$ channel then reads
\begin{align}
\hat{\sigma}_{gg}(z,Q_0)
 ={}&\sigma_0\, |C_t|^2 \bigg[\langle\bm{\Hh}_{2,gg}^{(0)}\otimes\mathbf{1}\rangle
 +\frac{\as}{4\pi}\big\langle
 \bm{\Hh}_{2,gg}^{(1)}\otimes\mathbf{1}
 +\bm{\Hh}_{2,gg}^{(0)}\otimes\bm{\cS}_{2,gg}^{(1)}
+\bm{\Hh}_{3,gg}^{(1)}\otimes\mathbf{1}\big\rangle \bigg].
 \label{eq:NLO_LP_expansion}
\end{align}
The remaining channels $qg \to qH$, $gq \to qH$ and $q\bar{q} \to gH$ start at NLO 
\begin{align}
\hat{\sigma}_{ij}(z,Q_0)
 ={}&\sigma_0\, |C_t|^2 \frac{\as}{4\pi}
\big\langle \bm{\Hh}_{3,ij}^{(1)}\otimes\mathbf{1}\big\rangle .
 \label{eq:NLO_LP_expansionij}
\end{align}
The fixed-order expansion is useful in the context of matching. The factorization theorem \eqref{eq:intro_factorization} and the associated resummation of logarithms are valid at leading power in an expansion around $Q_0/m_H \to 0$. To account for subleading terms, we add the difference between the full partonic cross section and \eqref{eq:NLO_LP_expansion} and \eqref{eq:NLO_LP_expansionij} to our resummed results. After this matching we obtain results which transition from resummed results at low $Q_0$ back to the NLO fixed order at larger values of $Q_0$, as discussed in Section~\ref{sec:matching2FO}. 

For the real-emission channels, we parameterize the scattering angle in the partonic center-of-mass frame by
\begin{align}
 y=\frac{1+\cos\theta}{2},\qquad 0\leq y\leq1\,.
 \label{eq:y_definition}
\end{align}
Due to the azimuthally symmetric cone constraint, the integral over the direction of the emission reduces to an integral over $y$. The emitted parton has rapidity $Y$ relative to the Higgs boson, with
\begin{align}
 Y=\frac12\ln\frac{\hat y}{1-\hat y},\qquad
 \hat y=\frac{y\bigl[y(1-z)+z\bigr]}{1-2(1-y)y(1-z)}.
 \label{eq:relative_rapidity}
\end{align}
The $\Theta_{\rm in}$ constraint on the hard emission can then be written in terms of these two variables as $\Theta_{\rm in}(z,y)=\theta(|Y|-Y_{\max})$. The distinction between $y$ and $\hat y$ matters: the hard kernels use the partonic center-of-mass angle $y$, while $\hat y$ is the angular variable in the frame in which the Higgs rapidity vanishes. The soft functions are sensitive to the measurement and must therefore be evaluated in the corresponding frame. We make their argument $\hat y(z,y)$ explicit below.

Up to NLO, the hard process involves at most three partons and each channel therefore only has a single color structure. We can thus take the color trace and reinstate the color structure later. We define the scalar hard coefficients through NLO as
\begin{align}
 \frac{\delta(y)+\delta(1-y)}{2} \langle\bm{\Hh}_{2,gg}(z)\rangle
 +\langle\bm{\Hh}_{3,gg}(z,y)\rangle
 &= C_{gg}(z,y)\Theta_{\rm in}(z,y),
 \label{eq:Hgg_match}\\
\langle\bm{\Hh}_{3,ij}(z,y)\rangle
 &= C_{ij}(z,y)\Theta_{\rm in}(z,y),
 \qquad ij=qg,gq,q\bar q. \nonumber
\end{align}
Here $C_{ij}=C_{ij}^{(0)}+\alpha_s C_{ij}^{(1)}/(4\pi)+\dots$ contains both the Born term and the NLO correction; the left-hand sides therefore involve the full hard coefficients, rather than their NLO corrections alone. In $C_{gg}(z,y)$, we have combined the real and virtual corrections. This is convenient since the individual parts suffer from collinear divergences which cancel in the combination. It is also compatible with \eqref{eq:reduced_factorized_cross_section} since in the collinear limit the three-leg soft function reduces to the Born-level one.

To obtain these kernels, we calculate the squared matrix elements of the contributions shown in Figure~\ref{fig:large_nc_mapping}. Their singular limits occur at $z\to1$ and $y\to0,1$. Initial-state collinear singularities are absorbed into the PDFs in the $\overline{\rm MS}$ scheme. The remaining soft poles of the angularly restricted bare hard functions cancel against the renormalization of the soft functions. The expressions below are the finite, renormalized hard kernels, with renormalization scale $\mu_h$ and $\alpha_s=\alpha_s(\mu_h)$.
\begin{align}
    C_{gg}(z,y) &=  \frac{\delta(1-z)\,\delta(y\bar{y})}{2} + \frac{\alpha_s}{4\pi} C_A \Bigg\{ \frac{\delta(1-z)\delta(y\bar{y})}{2} \left( 8 \zeta_2 + \frac{2}{C_A} \beta_0 \ln \frac{\mu_h^2}{\mu_f^2} \right) \nonumber \\
    &\qquad -4 \delta(y\bar{y}) \Bigg[  (z^2 - z + 1)^2 \left[ \frac{1}{1-z} \right]_+ \ln \frac{\mu_f^2}{m_H^2} +\frac{ (z^2 - z + 1)^2}{1-z}  \ln z \nonumber \\
    &\qquad\qquad\qquad\qquad\quad -2 \left[ \frac{\ln(1-z)}{1-z} \right]_+ +2 z(z^2 - z + 2) \ln(1-z) \Bigg] \nonumber \\
    &\qquad+   2 \delta(1-z) \left( \left[ \frac{\ln(y\bar{y})}{y\bar{y}} \right]_+ - \ln \frac{\mu_h^2}{m_H^2} \left[ \frac{1}{y\bar{y}} \right]_+ \right) \nonumber \\
    &\qquad +   4 \left[ \frac{1}{y\bar{y}} \right]_+ \left[ \left[ \frac{1}{1-z} \right]_+ - z(z^2 - z + 2) \right] - 4(1-z)^3 (y^2 - y + 2) \Bigg\} \,, \label{eq:Cgg} \\[11pt]
     C_{qg}(z,y) &= \frac{\alpha_s}{4\pi} C_F \Bigg\{ \delta(1-y) \left[ 2 [1+(1-z)^2] \left[\ln \left( \frac{(1-z)^2}{z} \right) - \ln \left( \frac{\mu_f^2}{m_H^2} \right)\right] + 2z^2 \right] \nonumber \\
    &\qquad \qquad \quad+ 2 \left[ [1+(1-z)^2] \left[ \frac{1}{1-y} \right]_+ - (1+y)(1-z)^2 \right] \Bigg\} \,, \label{eq:Cqg} \\
    C_{q\bar{q}}(z,y) &= \frac{\alpha_s}{4\pi} C_F \left\{ \frac{16}{3} (1-z)^3 [ y^2 + (1-y)^2 ] \right\} \,.\label{eq:Cqq}
\end{align}
The one-loop coefficient of the QCD $\beta$ function is 
\begin{align}
  \beta_0 = \frac{11}{3}C_A - \frac{4}{3}T_F n_f.
\end{align}
We use the shorthand notations
\begin{align}
    \bar{y} &= 1-y \, , & \delta(y\bar{y}) &= \delta(y) + \delta(\bar{y})\,,
\end{align} 
and work with symmetric plus distributions, which act on test functions $f(y)$ as
\begin{align}\label{eq:xibarxiplus}
  \int_0^1 \! dy  \left[\frac{1}{y \bar{y}}\right]_+ f(y) &=   \int_0^1 \! dy \left[\frac{1}{y \bar{y}} f(y)  -  \frac{1}{y} f(0) -   \frac{1}{\bar{y}} f(1)\right ] ,\\
   \int_0^1 \! dy \left[\frac{\ln(y \bar{y})}{y \bar{y}}\right]_+ f(y) &=   \int_0^1 \! dy \left[\frac{\ln(y \bar{y})}{y \bar{y}} f(y)  -  \frac{\ln y}{y} f(0) -   \frac{\ln \bar{y}}{\bar{y}} f(1)\right ] .
\end{align}
The kernel $C_{gq}(z,y)$ is related to $C_{qg}(z,y)$ via the symmetry relation 
\begin{align}
    C_{gq}(z,y) = C_{qg}(z, 1-y).
\end{align}
Replacing the cone constraint $\Theta_{\rm in}(z,y)$ with a measurement corresponding to an infrared safe observable, the above results become equivalent to the standard partonic cross section and we verified that~\eqref{eq:Cgg}--\eqref{eq:Cqq} agree with the expressions given in~\cite{Anastasiou:2002qz} for $\mu_h=\mu_f=m_H$. However, compared to \cite{Anastasiou:2002qz}, our gluon kernel contains the additional terms 
\begin{align}
    \Delta C_{gg} =  \frac{\alpha_s}{4\pi} C_A \delta(1-z) \Bigg\{ 2 \left[ \frac{\ln(y\bar{y})}{y\bar{y}} \right]_+ - 2 \ln \frac{\mu_h^2}{m_H^2} \left[ \frac{1}{y\bar{y}} \right]_+ \Bigg\} \,, \label{eq:deltasig}
\end{align}
which are proportional to $\delta(1-z)$ and therefore correspond to a soft gluon emission. For an infrared safe observable, the plus-distribution terms integrate to zero because the observable cannot be sensitive to the direction of the soft emission. However, we need to keep these terms, together with the angular restriction, because in our case, radiation inside the jet cones is part of the hard function, while emissions outside the cones are captured by the soft function. As a check of our implementation of the kernels, we compared the inclusive Higgs production cross section and its scale dependence against MCFM~\cite{Campbell:1999ah, Campbell:2011bn,Campbell:2019dru} in Appendix~\ref{app:crosschecks}.

\section{Resummation}
\label{sec:resummation}

\subsection{Evolution and matching ingredients}
To resum the NGLs to NLL accuracy, we evaluate the hard matching functions $\bm{\mathcal{H}}_{ij\to m}$ at their natural hard scale $\mu_h\sim m_H$, and evolve them down to a scale $\mu_s\sim Q_0$, where the soft functions $\bm{\mathcal{S}}_m$ are computed. The scale evolution of the hard functions is governed by the RG equations
\begin{align}
 \mu\frac{d}{d\mu}\Horig_{ij \to m}
 =-\sum_{l=2}^{m}\Horig_{ij\to l}\,
 \bm\Gamma_{lm}.
 \label{eq:hard_RGE_full}
\end{align}
This concerns the evolution in the scale $\mu$, for fixed factorization scale $\mu_f$. The evolution in $\mu_f$ gives rise to the usual PDF evolution, which is taken into account by evolving the PDFs to a scale $\mu_f\sim m_H$.

The formal solution to \eqref{eq:hard_RGE_full} can be written in terms of the evolution operator
\begin{align}
 \bm U_{lm}(\mu_h,\mu_s)
 =\left[\mathbf P\exp\left(\int_{\mu_s}^{\mu_h}
 \frac{d\mu}{\mu}\bm\Gamma\right)\right]_{lm}\,,
 \label{eq:evolution_operator}
\end{align}
so that
\begin{equation}
\Horig_m(\mu_s)  =\sum_{l=2}^{m} \Horig_l(\mu_h)\,\bm U_{lm}(\mu_h,\mu_s).
  \label{eq:Hlow}
\end{equation}
The matrix $\bm\Gamma$ acts on color and parton multiplicity, with angular integrations and intermediate multiplicity sums understood in products of evolution operators. The anomalous dimension is perturbatively expanded as $\bm\Gamma=\frac{\alpha_s}{4\pi}\bm\Gamma^{(1)}+\left(\frac{\alpha_s}{4\pi}\right)^2\bm\Gamma^{(2)}+\mathcal O(\alpha_s^3)$.

In the large-$N_c$ limit, the one-loop anomalous dimension $\bm{\Gamma}^{(1)}$ simplifies to a sum over color dipoles $[ij]$ formed by color-adjacent legs:
\begin{align}
 \bm\Gamma_{mn}^{(1)}=\sum_{[ij]}\left[
 \bm V_m^{ij}\delta_{m,n}+\bm R_m^{ij}\delta_{m,n-1}\right].
\end{align}
The diagonal virtual generator $\bm V_m^{ij}$ and off-diagonal real-emission generator $\bm R_m^{ij}$ are given by 
\begin{align}
 \bm R_m^{ij}=4N_cW_{ij}^q\Theta_{\rm in}(n_q),\qquad
 \bm V_m^{ij}=-4N_c\int[d^2\Omega_q]W_{ij}^q.
 \label{eq:LL_generator}
\end{align}

We use the normalization~\cite{Balsiger:2018ezi,Becher:2023vrh},
\begin{align}
 W_{ij}^q=\frac{n_i\cdot n_j}{(n_i\cdot n_q)(n_q\cdot n_j)},
 \qquad [d^2\Omega_q]=\frac{d\Omega_q}{4\pi},
 \qquad n_i=(1,\widehat{\boldsymbol n}_i).
 \label{eq:radiator_normalization}
\end{align}
The Heaviside function $\Theta_{\rm in}(n_q)$ prevents the emitted gluon from entering the gap region. 

At NLL accuracy, we incorporate the two-loop anomalous dimension $\bm{\Gamma}^{(2)}$, which describes double-virtual ($\bm{v}_m^{ij}$), real-virtual ($\bm{r}_m^{ij}$), and double-real ($\bm{d}_m^{ij}$) contributions~\cite{Caron-Huot:2015bja,Becher:2021urs}:
\begin{align}
 \bm\Gamma_{mn}^{(2)}=\sum_{[ij]}\left[
 \bm v_m^{ij}\delta_{m,n}+\bm r_m^{ij}\delta_{m,n-1}
 +\bm d_m^{ij}\delta_{m,n-2}\right].
\end{align}
The matrix structures of $\bm\Gamma^{(1)}$ and $\bm\Gamma^{(2)}$ are block-triangular, reflecting the transition of parton configurations to higher multiplicities. The two-loop kernel and the one-loop hard and soft functions depend on the renormalization scheme. For our computation we adopt the Lorentz subtraction ($\overline{\mathrm{LS}}$) scheme~\cite{Caron-Huot:2015bja,Becher:2026zon}. The large-$N_c$ expression for the two-loop anomalous dimension in this scheme can be found in \cite{Becher:2026zon}.

To solve the RG equation, it is convenient to introduce the evolution time
\begin{align}
 t(\mu_h,\mu_s)=\frac{1}{2\beta_0}
 \ln\frac{\alpha_s(\mu_s)}{\alpha_s(\mu_h)}\, .
 \label{eq:evolution_time}
\end{align}
The LL evolution then takes the simple form $\bm U^{\rm LL}(t_0,t)=\exp[(t-t_0)\bm\Gamma^{(1)}]$ and is solved iteratively in small time steps by using a parton shower Monte Carlo. 

The NLL correction $\Delta\bm U_{kl}(t_0,t)$ involves a single insertion of the two-loop anomalous dimension into the LL shower history~\cite{Becher:2023vrh}:
\begin{align}
 \Delta\bm U_{kl}(t_0,t)
 =\int_{t_0}^{t}dt'\,
 \bm U^{\rm LL}_{kk'}(t_0,t')\,
 \frac{\alpha_s(t')}{4\pi}
 \left(\bm\Gamma^{(2)}_{k'l'}-
 \frac{\beta_1}{\beta_0}\bm\Gamma^{(1)}_{k'l'}\right)
 \bm U^{\rm LL}_{l'l}(t',t)\, .
 \label{eq:NLL_insertion}
\end{align}
 The term proportional to $\beta_1$ accounts for the two-loop running of the coupling $\alpha_s$. We start the evolution at $\mu_h$ which corresponds to $t_0=0$. 

To write the resummed result in a compact form, it is helpful to combine the evolution with the soft functions at the low scale. We define
\begin{align}
 \bm{\mathcal S}^{(n),\rm LL}_{m,ij}(t)
 &=\sum_{k\geq m}\bm U^{\rm LL}_{mk}(0,t)\,\widehat\otimes\,
 \bm{\mathcal S}^{(n)}_{k,ij}(\mu_s),\qquad n=0,1,
 \label{eq:evolved_soft_definition}\\
 \Delta\bm{\mathcal S}^{(0),\rm NLL}_{m,ij}(t)
 &=\sum_{k\geq m}\Delta\bm U_{mk}(0,t)\,\widehat\otimes\,\mathbf 1.
 \label{eq:inserted_soft_definition}
\end{align}
The convolution $\widehat\otimes$ integrates over the additional directions generated by evolution and we used $\bm{\mathcal S}^{(0)}_{k,ij}(\mu_s)=\mathbf 1$. In this notation the NLL resummed partonic cross section for the $gg$-channel contains five terms:
\begin{align}
 \hat{\sigma}_{gg}^{\rm NLL}(z,Q_0)
 ={}&  \left\langle\bm{\cH}^{(0)}_{2,gg}\otimes\bm{\mathcal S}^{(0),\rm LL}_{2,gg}(t)\right\rangle\nonumber\\
 &+\frac{\alpha_s(\mu_h)}{4\pi}\left[
 \left\langle\bm{\mathcal H}^{(1)}_{2,gg}\otimes\bm{\mathcal S}^{(0),\rm LL}_{2,gg}(t)\right\rangle+
 \left\langle\bm{\mathcal H}^{(1)}_{3,gg}\otimes\bm{\mathcal S}^{(0),\rm LL}_{3,gg}(t)\right\rangle\right]\nonumber\\
 &+\frac{\alpha_s(\mu_s)}{4\pi}
 \left\langle\bm{\mathcal H}^{(0)}_{2,gg}\otimes\bm{\mathcal S}^{(1),\rm LL}_{2,gg}(t)\right\rangle+
 \left\langle\bm{\mathcal H}^{(0)}_{2,gg}\otimes\Delta\bm{\mathcal S}^{(0),\rm NLL}_{2,gg}(t)\right\rangle \,.
 \label{eq:NLL_master}
\end{align}
The first term represents the LL resummation. The second and third terms account for the NLO hard matching corrections to the two- and three-parton configurations. The fourth term is the one-loop soft matching correction, and the fifth term represents the insertion of the two-loop anomalous dimension. For the remaining partonic channels, $ij\neq gg$, LL evolution is sufficient since they only enter at NLO:
\begin{align}
 \hat{\sigma}_{ij}^{\rm NLL}(z,Q_0)
 ={}& \frac{\alpha_s(\mu_h)}{4\pi}
 \left\langle\bm{\mathcal H}^{(1)}_{3,ij}\otimes\bm{\mathcal S}^{(0),\rm LL}_{3,ij}(t)\right\rangle .
 \label{eq:NLL_masterij}
\end{align}
We can simplify the equations further since the NLL hard functions contain at most three colored partons and therefore involve only a single color structure. In this case also the soft function color structure is trivial and we can replace
\begin{align}
 \bm{\mathcal S}_{2,gg}(t) &= \mathcal {S}_{2,gg}(t)\, \bm{1}\,, & \bm{\mathcal S}_{3,ij}(t) &= \mathcal {S}_{3,ij}(t)\, \bm{1}\,.
\end{align}
Scalar soft functions are written in regular typeface, in contrast to the bold symbols used for their color-matrix counterparts. Inserting this into \eqref{eq:NLL_master} and \eqref{eq:NLL_masterij}, we obtain the cross section directly in terms of the hard kernels and scalar soft functions. The $gg$ channel reads
\begin{align}
 \hat{\sigma}_{gg}^{\rm NLL}(z,Q_0)
 ={}& \sigma_0 |C_t|^2 \int_0^1 \!dy \,\Theta_{\rm in}(z,y) \bigg\{ C^{(0)}_{gg}(z,y) {\mathcal S}^{(0),\rm LL}_{2,gg}(t)+\frac{\alpha_s(\mu_h)}{4\pi}
 C^{(1)}_{gg}(z,y) {\mathcal S}^{(0),\rm LL}_{3,gg}(t,\hat{y})\nonumber\\
 &\quad\quad +\frac{\alpha_s(\mu_s)}{4\pi} C^{(0)}_{gg}(z,y)
 {\mathcal S}^{(1),\rm LL}_{2,gg}(t)+
C^{(0)}_{gg}(z,y) \Delta {\mathcal S}^{(0),\rm NLL}_{2,gg}(t) \bigg\} \,.
 \label{eq:NLL_master2}
\end{align}
In contrast to \eqref{eq:NLL_master}, this equation only contains four terms: the virtual corrections have been absorbed into $C^{(1)}_{gg}(z,y)$, see~\eqref{eq:Hgg_match}, and the soft function ${\mathcal S}^{(0),\rm LL}_{3,gg}(t,\hat{y})$ reduces to the two-parton result for $y\to 0,1$. The remaining channels start at NLO and take the form
\begin{align}
 \hat{\sigma}_{ij}^{\rm NLL}(z,Q_0)
 ={}&  \sigma_0 |C_t|^2 \int_0^1 \!dy \,\Theta_{\rm in}(z,y) \frac{\alpha_s(\mu_h)}{4\pi}
 C^{(1)}_{ij}(z,y) {\mathcal S}^{(0),\rm LL}_{3,ij}(t,\hat{y}).
 \label{eq:NLL_masterij2}
\end{align}

\subsection{Universal soft evolution for Higgs and Drell--Yan production}

The soft matrix elements $\bm{\mathcal S}_{ij\to m}$ and their evolution depend on the geometry of the veto region and are sensitive to the color charges and directions of the partons sourcing the radiation. Other than this, they are universal and independent of the underlying hard-scattering process. For partonic channels which are present both in Drell--Yan and in Higgs production, we can directly use the soft functions obtained in \cite{Becher:2023vrh} for the Drell--Yan case. This includes ${\mathcal S}_{3,qg}(t,\hat y)$ associated with $qg \to q \,Z/H $, as well as ${\mathcal S}_{3,q\bar q}(t,\hat y)$ from $q\bar{q} \to g\, Z/H $. Since these channels only arise at NLO, we only need the corresponding soft functions at LL accuracy.

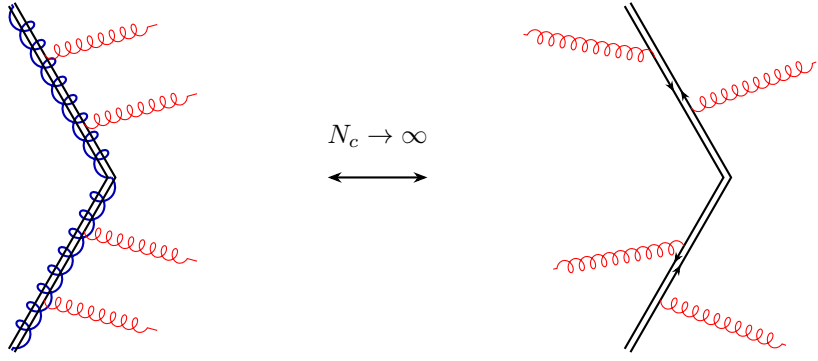
\begin{figure}[t!]
    \centering
    \scalebox{1.0}{
    \begin{tikzpicture}[scale=2.2]
      \begin{feynman}
        \vertex (originL) at (3.5, 0); 
        \vertex (g_topL) at ($(originL) + (120:1.2)$);   
        \vertex (g_bottomL) at ($(originL) + (240:1.2)$); 
    
        \diagram* {
          (originL) -- [thick, blue!70!black, decorate, decoration={coil, aspect=0.85, amplitude=0.13cm, segment length=0.28cm}] (g_topL),
          (originL) -- [thick, blue!70!black, decorate, decoration={coil, aspect=0.85, amplitude=0.13cm, segment length=0.28cm}] (g_bottomL),
        };
    
        \def\dist{0.02cm}
        \coordinate (v_outerL) at ($(originL) + (0:0.023cm)$); 
        \coordinate (v_innerL) at ($(originL) + (180:0.023cm)$); 
        \draw [thick, black, line join=miter] ($(g_topL) + (30:\dist)$) -- (v_outerL) -- ($(g_bottomL) + (330:\dist)$);
        \draw [thick, black, line join=miter] ($(g_topL) + (210:\dist)$) -- (v_innerL) -- ($(g_bottomL) + (150:\dist)$);
    
        \coordinate (st1L) at ($($(originL)!0.3!(g_topL)$) + (30:\dist)$);
        \coordinate (st2L) at ($($(originL)!0.7!(g_topL)$) + (30:\dist)$);
        \coordinate (sb1L) at ($($(originL)!0.3!(g_bottomL)$) + (330:\dist)$);
        \coordinate (sb2L) at ($($(originL)!0.7!(g_bottomL)$) + (330:\dist)$);
    
        \vertex (et1L) at ($(st1L) + (15:0.7)$);
        \vertex (et2L) at ($(st2L) + (15:0.7)$);
        \vertex (eb1L) at ($(sb1L) + (-15:0.7)$);
        \vertex (eb2L) at ($(sb2L) + (-15:0.7)$);
    
        \diagram* {
          (st1L) -- [gluon, red, small] (et1L),
          (st2L) -- [gluon, red, small] (et2L),
          (sb1L) -- [gluon, red, small] (eb1L),
          (sb2L) -- [gluon, red, small] (eb2L),
        };
    
    
        \draw [{Stealth[scale=0.8]}-{Stealth[scale=0.8]}, thick] (4.8, 0) -- (5.4, 0);
        \node[font=\small] at (5.1, 0.25) {$N_c \to \infty$ };
    
        \vertex (originR) at (7.2, 0); 
        \vertex (g_topR) at ($(originR) + (120:1.2)$);   
        \vertex (g_bottomR) at ($(originR) + (240:1.2)$); 
    
        \coordinate (v_outerR) at ($(originR) + (0:0.023cm)$);   
        \coordinate (v_innerR) at ($(originR) + (180:0.023cm)$); 
    
        \coordinate (top_outer_endR) at ($(g_topR) + (30:\dist)$);
        \coordinate (top_inner_endR) at ($(g_topR) + (210:\dist)$);
        \coordinate (bot_outer_endR) at ($(g_bottomR) + (330:\dist)$);
        \coordinate (bot_inner_endR) at ($(g_bottomR) + (150:\dist)$);
    
        \tikzset{
          colorflow_arrow/.style={
            thick, black, decoration={markings, 
            mark=at position 0.25 with {\arrow{Stealth[length=4pt]}},
            mark=at position 0.75 with {\arrow{Stealth[length=4pt]}}},
            postaction={decorate}
          }
        }
    
        \draw [colorflow_arrow] (top_inner_endR) -- (v_innerR) -- (bot_inner_endR);
        \draw [colorflow_arrow] (bot_outer_endR) -- (v_outerR) -- (top_outer_endR);
    
        \coordinate (st_inR)  at ($(v_innerR)!0.7!(top_inner_endR)$);
        \coordinate (st_outR) at ($(v_outerR)!0.4!(top_outer_endR)$);
        \coordinate (sb_inR)  at ($(v_innerR)!0.4!(bot_inner_endR)$);
        \coordinate (sb_outR) at ($(v_outerR)!0.7!(bot_outer_endR)$);
    
        \vertex (et_inR)  at ($(st_inR) + (170:0.8)$);
        \vertex (et_outR) at ($(st_outR) + (20:0.8)$);
        \vertex (eb_inR)  at ($(sb_inR) + (190:0.8)$);
        \vertex (eb_outR) at ($(sb_outR) + (340:0.8)$);
    
        \diagram* {
          (st_inR)  -- [gluon, red, small] (et_inR),
          (st_outR) -- [gluon, red, small] (et_outR),
          (sb_inR)  -- [gluon, red, small] (eb_inR),
          (sb_outR) -- [gluon, red, small] (eb_outR),
        };
      \end{feynman}
    \end{tikzpicture}
    }
    \caption{Planar color structure of the two incoming gluons. Their color connections form two independent back-to-back dipoles. The gluon-initiated soft evolution is obtained by combining the evolution of these two dipoles.}
    \label{fig:nc_soft_radiation}
\end{figure}

\begin{figure}[t!]
    \centering
    \begin{subfigure}{\textwidth}
    \centering
        \includegraphics[height=4.0cm]{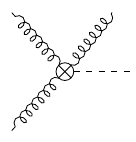}
        \includegraphics[height=4.0cm]{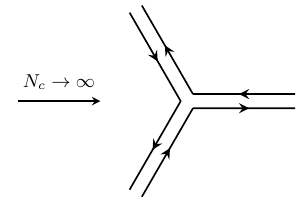}
    \hfill
    \subcaption[]{\(gg \to gH\) channel}
    \end{subfigure}\\[12pt]
    \begin{subfigure}{\textwidth}
    \centering
        \includegraphics[height=4.0cm]{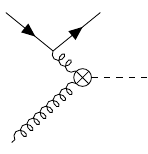}
        \includegraphics[height=4.0cm]{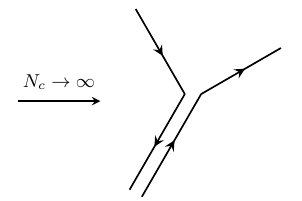}
    \hfill
    \subcaption[]{\(qg \to qH\) channel}
    \end{subfigure}\\[12pt]
        \begin{subfigure}{\textwidth}
    \centering
        \includegraphics[height=4.0cm]{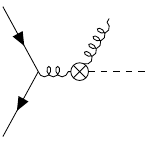}
        \includegraphics[height=4.0cm]{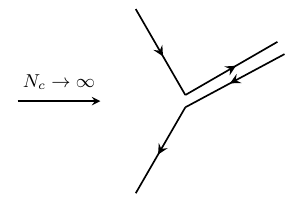}
    \hfill
    \subcaption[]{\(q\bar{q} \to gH\) channel}
    \end{subfigure}
     \caption{Example diagrams for each real-emission partonic channel, together with the associated large-$N_c$ color structure. The circle-times symbol indicates the effective vertex after integrating out the top quark. The soft functions only resolve the color structure of the hard functions and the directions of the three partons.}
    \label{fig:large_nc_mapping}
\end{figure}

In the large-$N_c$ limit, we obtain additional relations since the color structure of an amplitude reduces to a set of dipoles, see Figures~\ref{fig:nc_soft_radiation} and~\ref{fig:large_nc_mapping}. For fixed angular constraints, these dipoles evolve independently and we can multiply results for individual dipoles to obtain more complicated soft functions. 
For a given color flow, we end up with a scalar soft function. The simplest example of a large-$N_c$ relation is the Born channel $gg\to H$, in which each incoming gluon carries two planar color lines. Figure~\ref{fig:nc_soft_radiation} shows that these form two back-to-back dipoles, whereas $q\bar q\to Z$ has one. Consequently,
\begin{align}
 \mathcal S_{2,gg}(t)
 =\left[\mathcal S_{2,q\bar q}(t)\right]^2\,,
 \label{eq:map_gg}
\end{align}
which is a statement about two independent radiation systems in the large-$N_c$ limit, not a replacement of a quark color factor by a gluon color factor. Each factor includes the full non-global branching of its dipole. We can expand the relation \eqref{eq:map_gg} to NLL, to obtain relations up to first order in $\alpha_s$. They read
\begin{align}
 \mathcal S_{2,gg}^{(0),\rm LL}(t)
 &=\left[\mathcal S_{2,q\bar q}^{(0),\rm LL}(t)\right]^2, \\
  \mathcal S_{2,gg}^{(1),\rm LL}(t)
 &=2\,\mathcal S_{2,q\bar q}^{(0),\rm LL}(t)\,
 \mathcal S_{2,q\bar q}^{(1),\rm LL}(t),
 \label{eq:map_soft_matching}\\
 \Delta\mathcal S_{2,gg}^{(0),\rm NLL}(t)
 &=2\,\mathcal S_{2,q\bar q}^{(0),\rm LL}(t)\,
 \Delta\mathcal S_{2,q\bar q}^{(0),\rm NLL}(t).
 \label{eq:map_soft_insertion}
\end{align}
The latter two relations simply express that we can add a one-loop correction to the soft function or the evolution in either one of the two dipoles.

The final partonic process not present in the Drell--Yan case at NLO is $gg \to H g$. Diagram (a) in Figure~\ref{fig:large_nc_mapping} shows that its color structure splits into an incoming $q\bar{q}$ dipole, multiplied by the same dipole structure that is present in the $q\bar{q} \to H g$ channel, shown in Diagram (c). This gives the large-$N_c$ relation
\begin{align}
 \mathcal S_{3,gg}^{(0),\rm LL}(t, \hat y)
 =\mathcal S_{2,q\bar q}^{(0),\rm LL}(t)\,
 \mathcal S_{3,q\bar q}^{(0),\rm LL}(t,\hat y).
 \label{eq:map_ggg}
\end{align}

Together, the above relations supply every soft ingredient in~\eqref{eq:NLL_master2} and~\eqref{eq:NLL_masterij2} from the existing Drell--Yan calculation in \cite{Becher:2023vrh}, which computed the soft functions $\mathcal S_{3,ij}^{(0),\rm LL}(t,\hat y)$ on a grid of $\hat{y}$- and $t$-values. Convolving the Higgs kernels $C_{ij}$ with interpolated soft functions then leads to the numerical results presented in the next section.

\subsection{Matching to fixed order}
\label{sec:matching2FO}
The factorization theorem in~\eqref{eq:intro_factorization} and the resummation described above are valid for $Q_0\ll m_H$.
However, for large values of $Q_0 \sim m_H$, the fixed-order cross section provides the appropriate prediction and the factorization theorem is no longer valid.
In order to obtain a prediction that is reliable for all values of $Q_0$, we match the resummed and the fixed-order cross sections:
\begin{equation}
  \sigma^{\text{NLL+NLO}}(Q_0) = \sigma^{\text{NLL}}(Q_0)
  + \Big[\, \sigma^{\text{NLO}}(Q_0)
  - \sigma^{\text{NLL}}(Q_0)\big|_{\mathcal{O}(\alpha_s)} \,\Big]. 
  \label{eq:matched}
\end{equation}
The first term in the sum corresponds to the resummed prediction at NLL which is accurate for $Q_0 \ll m_H$. The term in brackets is the nonsingular matching correction. It is obtained by subtracting from the fixed-order cross section its singular limit, which is the same as evaluating the factorization theorem without evolution. This removes the soft contributions from real emissions inside the gap that are already contained in $\sigma^{\text{NLL}}$ and would otherwise be 
double counted.
What remains is power suppressed for $Q_0 \ll m_H$, so that the resummed prediction is left unaffected, while for $Q_0 \sim m_H$ it restores the full fixed-order result.

In addition to the matching, we need to switch off the resummation in the fixed-order region as
for $Q_0 \sim m_H$ the
logarithms $\ln(m_H/Q_0)$ are no longer large, and resumming them introduces spurious
higher-order terms. 
Profile
functions~\cite{Abbate:2010xh} achieve this smoothly by making the soft scale a function
of $Q_0$. Following~\cite{Balsiger:2019tne}, we use
\begin{equation}
  \mu_s(Q_0) = \frac{x_s\, Q_0}
  {\big(1 - Q_0/m_H\big)^{n} + x_s\, Q_0/\mu_h} \,,
  \label{eq:profile}
\end{equation}
which interpolates between the two regimes. Our standard choice is $n=4$, which sharply turns off the resummation at higher $Q_0$ values, but we will also show results for $n=2$ for comparison. For $Q_0 \ll m_H$ the denominator
approaches unity and the canonical choice $\mu_s = x_s Q_0$ is recovered, leaving the
resummation unaffected. As $Q_0$ approaches $m_H$, one finds $\mu_s \to \mu_h$ and
therefore $t(\mu_h,\mu_s) \to 0$, so that the evolution is switched off and
~\eqref{eq:matched} reduces to the fixed-order prediction. The soft scale is varied
in the range $1/2 \le x_s \le 2$ around the central value $x_s = 1$. 

We refer to the matched cross section in~\eqref{eq:matched}, evaluated with the
profile scale of~\eqref{eq:profile}, as the NLL+NLO prediction, since it is accurate
at NLL in the resummation region and reduces to the NLO cross section in the fixed-order
region.

\section{Phenomenological predictions}
\label{sec:numerical_results}
In this section, we present numerical results for the gap fraction $R(Q_0)$ in the gluon fusion process.
We compare formally equivalent expansion schemes, study the scale uncertainties and the effect of the profile-function choice.

\subsection{Expansion schemes}
There are different ways to define the gap fraction perturbatively, which agree at the working order but differ in how the ratio of the vetoed and inclusive cross sections is expanded, and hence in their residual scheme dependence.
To illustrate this, we consider the fixed-order expansion of the gap fraction at $\mathcal{O}(\alpha_s)$, noting that there is no radiation at LO, such that $\sigma^{(0)}(Q_0) = \sigma_{\text{incl}}^{(0)}$.
One approach is to expand the individual cross sections in the numerator and denominator separately,
\begin{align}
    R_{\text{unexp}}(Q_0) &= \frac{\sigma_{\text{incl}}^{(0)} + \alpha_s \sigma^{(1)}(Q_0) + \mathcal{O}(\alpha_s^2)}{\sigma_{\text{incl}}^{(0)} + \alpha_s \sigma_{\text{incl}}^{(1)} + \mathcal{O}(\alpha_s^2)} \,,
\end{align}
which we refer to as the unexpanded scheme. Note that  $\sigma_{\text{incl}}^{(1)}$ refers to the coefficient of $\alpha_s$, without the $1/(4\pi)$ factor present in \eqref{eq:hard_soft_expansions}. Here the ratio structure is preserved: the vetoed and the inclusive cross section are each evaluated to the working order, and their ratio is taken without further expansion.
However, common factors such as $C_t^2$ and the Born luminosity cancel only at leading order, so residual inclusive corrections and $\mu_f$ dependence remain part of the ratio and can lead to larger scale uncertainties.

In the expanded scheme, we instead expand the gap fraction directly, rather than the individual cross sections,
\begin{align}
    R_{\text{exp}}(Q_0) &= 1 + \alpha_s \left( \frac{ \sigma^{(1)}(Q_0) - \sigma_{\text{incl}}^{(1)}}{\sigma_{\text{incl}}^{(0)}}  \right) + \mathcal{O}(\alpha_s^2) \,.
\end{align}
Here the common factors cancel exactly, so the deviation from unity is a pure $\mathcal{O}(\alpha_s)$ effect.
This generally leads to a smaller scale dependence. Whether the reduction is physical is not obvious a priori: the cancellation itself is genuine, since the gap fraction is independent of overall factors such as $|C_t|^2$.

The analytic continuation of form factors from spacelike to timelike momentum transfer generates $\pi^2$-enhanced higher-order terms which can be predicted to all orders with resummation methods since they originate from logarithms~\cite{Parisi:1979xd,Sterman:1986aj,Magnea:1990zb}. It was pointed out in~\cite{Ahrens:2008qu} that the scalar gluon form factor $C_S(Q^2,\mu^2)$ has a well-behaved perturbative expansion for spacelike argument $Q^2>0$, while
its timelike expansion converges poorly. In the factorization theorem \eqref{eq:reduced_factorized_cross_section} the timelike gluon form factor defines the $gg\to 0$ hard function
\begin{equation}
\langle \bm{\Hh}_{2,gg}\rangle = \big|C_S(-\hat{s},\mu_h^2)\big|^2 = 1 + \frac{\alpha_s}{4\pi}\,C_A
  \left( -2\ln^2\frac{\hat{s}}{\mu_h^2} + \frac{7\pi^2}{3} \right)
  + \mathcal{O}(\alpha_s^2) \,,
  \label{eq:CS}
\end{equation}
so that the large $\pi^2$-correction is present in our result. The work~\cite{Ahrens:2008qu,Ahrens:2009cxz} has resummed the $\pi^2$ terms, using RG-evolution from a timelike to a spacelike scale choice and has demonstrated that this improves the convergence of the perturbative expansion of the inclusive Higgs production cross section. Assuming that the higher-multiplicity hard functions suffer from similar corrections, it makes sense to factor them out both in the numerator and denominator of \eqref{eq:gap_fraction}, before forming the
gap fraction, which in practice amounts to subtracting its $\mathcal{O}(\alpha_s)$
expansion from $\sigma_{\text{incl}}$ and from the hard correction in the resummed vetoed cross section. We denote the resulting scheme by $R_{\pi^2}(Q_0)$. Since the same correction is removed from numerator and denominator, it cancels in the expanded scheme and modifies only the unexpanded one.

We emphasize that all three schemes present valid choices and that their difference is formally beyond our working order. The differences can nevertheless be sizable, and we study them in detail in the remainder of this section.

\subsection{Results}
\label{sec:pheno_predictions}

\begin{figure}
    \centering
    \includegraphics[width=0.6\textwidth]{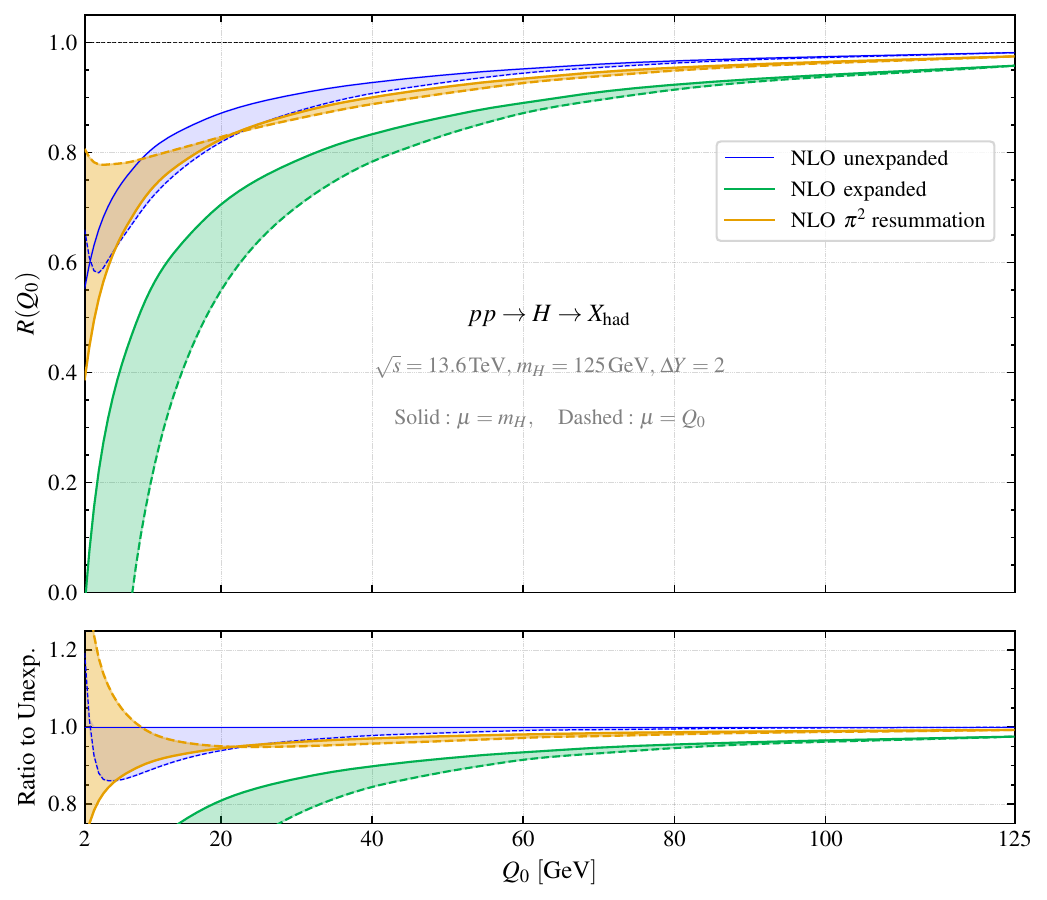}
    \caption{Fixed-order gap fraction $R(Q_0)$ for $pp \to H \to X_{\rm had}$ for the three different expansion schemes: unexpanded, expanded and $\pi^2$ resummation.}
    \label{fig:gap_fraction_NLO}
\end{figure}

For our numerical predictions, we consider $pp$ collisions at $\sqrt{s}=13.6\,\text{TeV}$ with a rapidity gap $\Delta Y = 2$. Parton distributions are evaluated using the \texttt{NNPDF40\_nlo\_as\_01180} set~\cite{NNPDF:2021njg}. For the Fermi constant we use $G_F = 1.16639 \times 10^{-5}\,\text{GeV}^{-2}$~\cite{ParticleDataGroup:2026mpi}, and the vacuum expectation value is fixed via $v = (\sqrt{2}\,G_F)^{-1/2}$.

Figure~\ref{fig:gap_fraction_NLO} shows the fixed-order gap fraction at NLO for the
unexpanded scheme in blue, the expanded scheme in green and the $\pi^2$ resummation in
orange. Solid lines correspond to the scale choice $\mu=m_H$, dashed lines to
$\mu=Q_0$. As commonly done, we take the envelope of the two scale choices as our
fixed-order uncertainty.
Physically, the gap fraction should vanish in the limit $Q_0\to0$, since this
corresponds to allowing no energy in the gap. The fixed-order predictions fail to
reproduce this behavior: the gap fraction diverges to $\pm \infty$, depending on the scale choice. This simply reflects the growth of the
logarithms $\ln(m_H/Q_0)$ towards small $Q_0$, which spoils the convergence of the
perturbative series. These logarithms must be resummed to obtain a reliable
phenomenological prediction and to restore the correct behavior of the gap fraction
as $Q_0\to0$.
We further find that the fixed-order prediction depends strongly on the expansion
scheme introduced in the previous section. This discrepancy is significantly larger
than the scale variation within any individual scheme and is therefore not captured
by the standard uncertainty estimate. Even at large $Q_0$, where the fixed-order
prediction is valid, the scale uncertainty bands of the different schemes do not
overlap.
This shows that the scheme dependence is unrelated to the breakdown of the
perturbative expansion at small $Q_0$.

\begin{figure}
    \centering
    \begin{subfigure}{0.48\textwidth}
        \includegraphics[width=\linewidth]{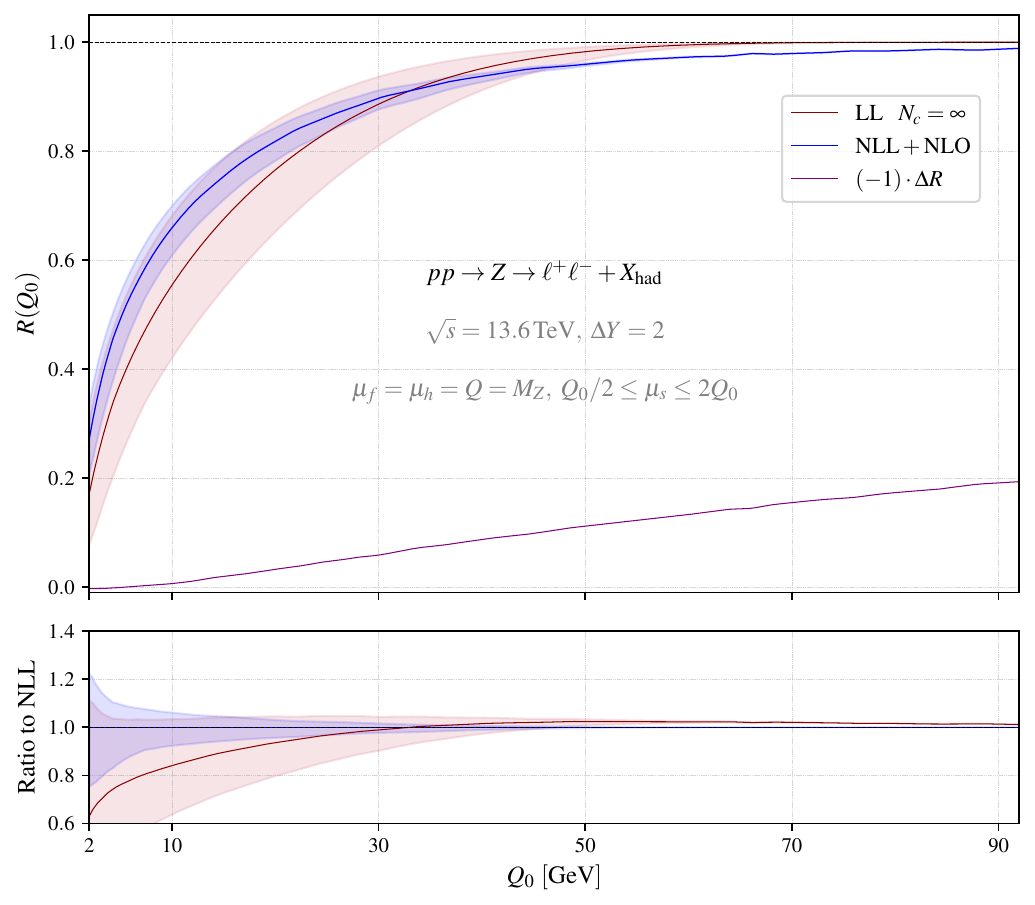}
        \caption{Drell--Yan process }
    \end{subfigure}
    \hfill
    \begin{subfigure}{0.48\textwidth}
        \includegraphics[width=\linewidth]{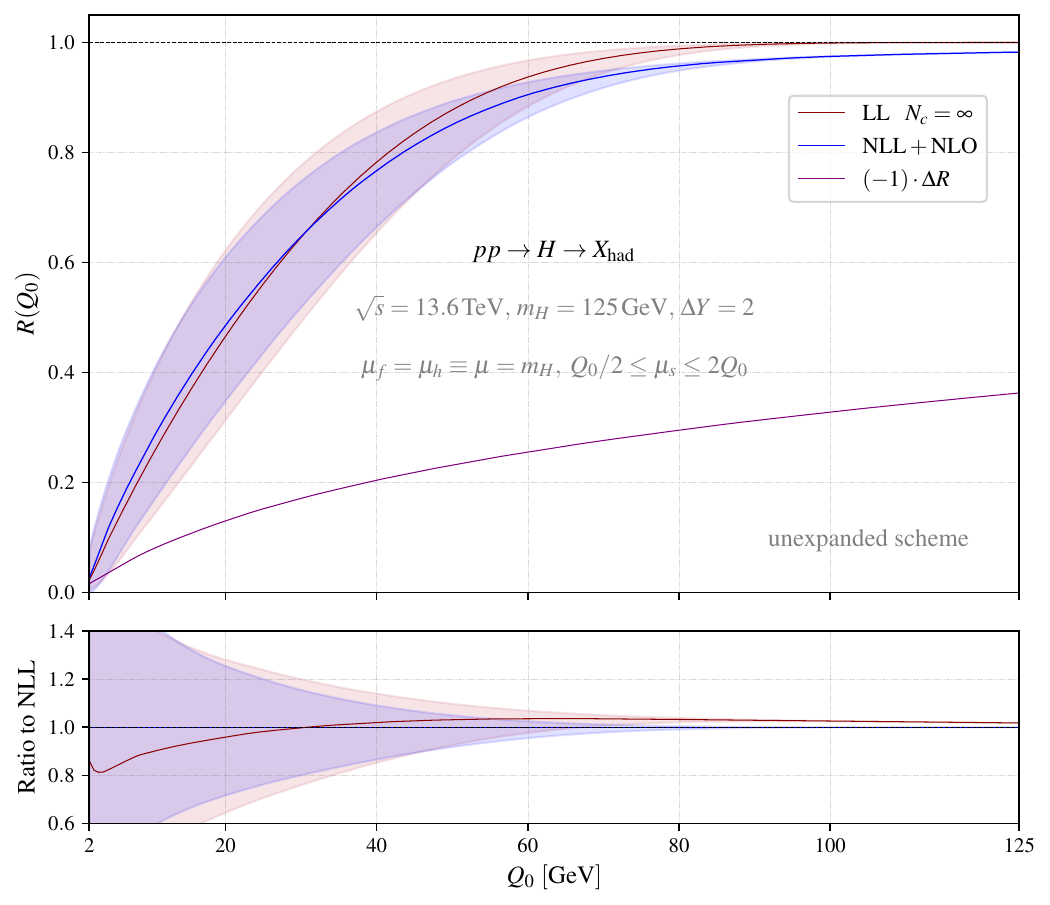}
        \caption{Higgs production}
    \end{subfigure}
    \caption{Resummed gap fraction $R(Q_0)$ for (a) Drell--Yan and (b) Higgs production at $\sqrt{s} = 13.6\,\mathrm{TeV}$ with scale uncertainty bands $Q_0/2 \le \mu_s \le 2Q_0$.}
    \label{fig:gap_fraction_ggH_DY}
\end{figure}

In Figure~\ref{fig:gap_fraction_ggH_DY} we show the resummed gap fraction for Higgs
production and the results for the Drell--Yan process from reference \cite{Becher:2023vrh}, but with updated soft evolution and PDFs. The LL prediction is shown in red and the NLL+NLO prediction
in blue, while the purple line gives the nonsingular matching contribution. The uncertainty bands
are obtained by varying the soft scale $\mu_s$ by a factor of two around its central
value $\mu_s = Q_0$. For large values of $Q_0$ the resummation is turned off using the
profile function of Section~\ref{sec:matching2FO}, and we recover the fixed-order prediction.

In general, we observe a reduction of the perturbative uncertainties from LL to
NLL+NLO. For Higgs production, however, the uncertainties remain sizable even at
NLL+NLO, particularly in comparison to the Drell--Yan case, where they decrease by a
factor of two between LL and NLL+NLO. The matching correction is also considerably
larger in Higgs production: for Drell--Yan it stays below 20\% throughout, whereas for
Higgs production it exceeds 20\% already at $Q_0 = 40\,\rm GeV$ and grows to $\sim 35\%$
at $Q_0 = 125\,\rm GeV$. Both observations are expected, since the $\cO(\as)$ corrections
in gluon fusion are known to be large. This is well established for the inclusive cross
section~\cite{DAWSON1991283,Djouadi:1991tka,Spira:1995rr} and is also seen in resummed
cross sections~\cite{Ebert:2017uel,Re:2021con,Cal:2023mib}. In addition, the
initial-state gluons enhance the effects of NGLs in gluon fusion compared to Drell--Yan.

\begin{figure}
    \centering
    \begin{subfigure}{0.48\textwidth}
        \includegraphics[width=\linewidth]{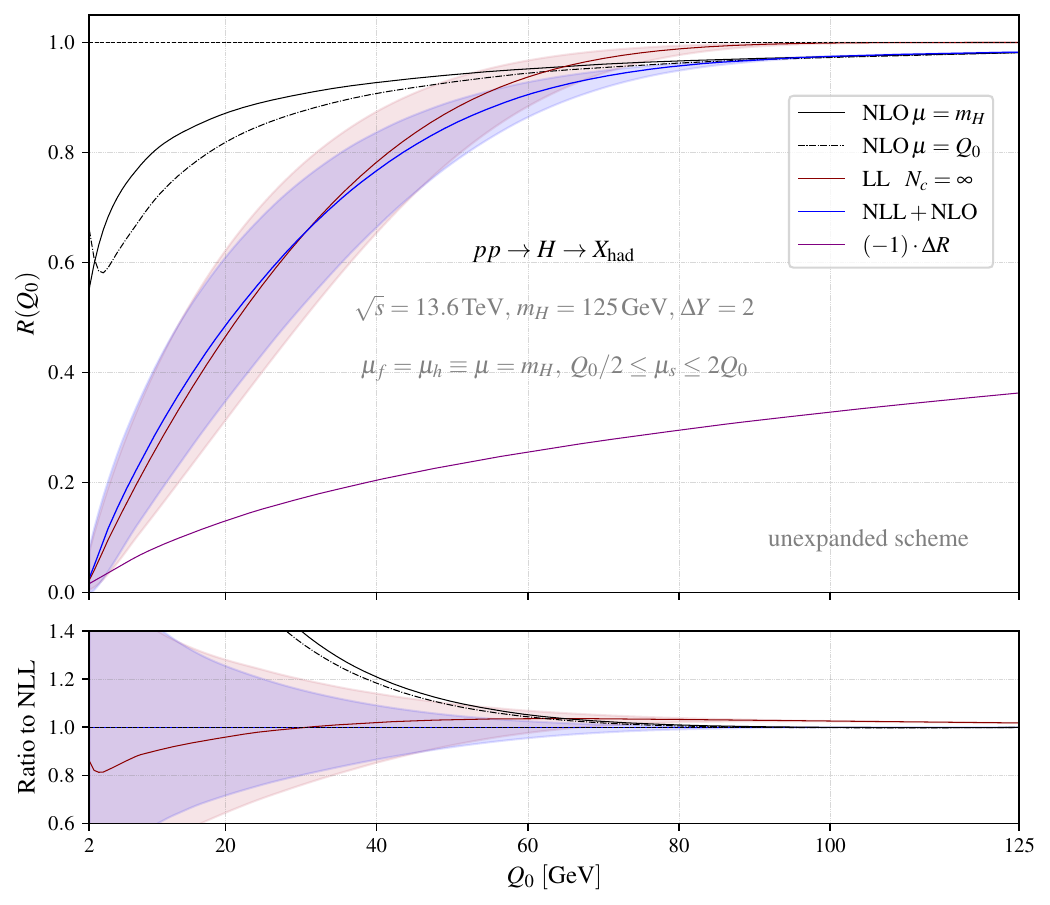}
        \caption{Unexpanded scheme}
    \end{subfigure}
    \hfill
    \begin{subfigure}{0.48\textwidth}
        \includegraphics[width=\linewidth]{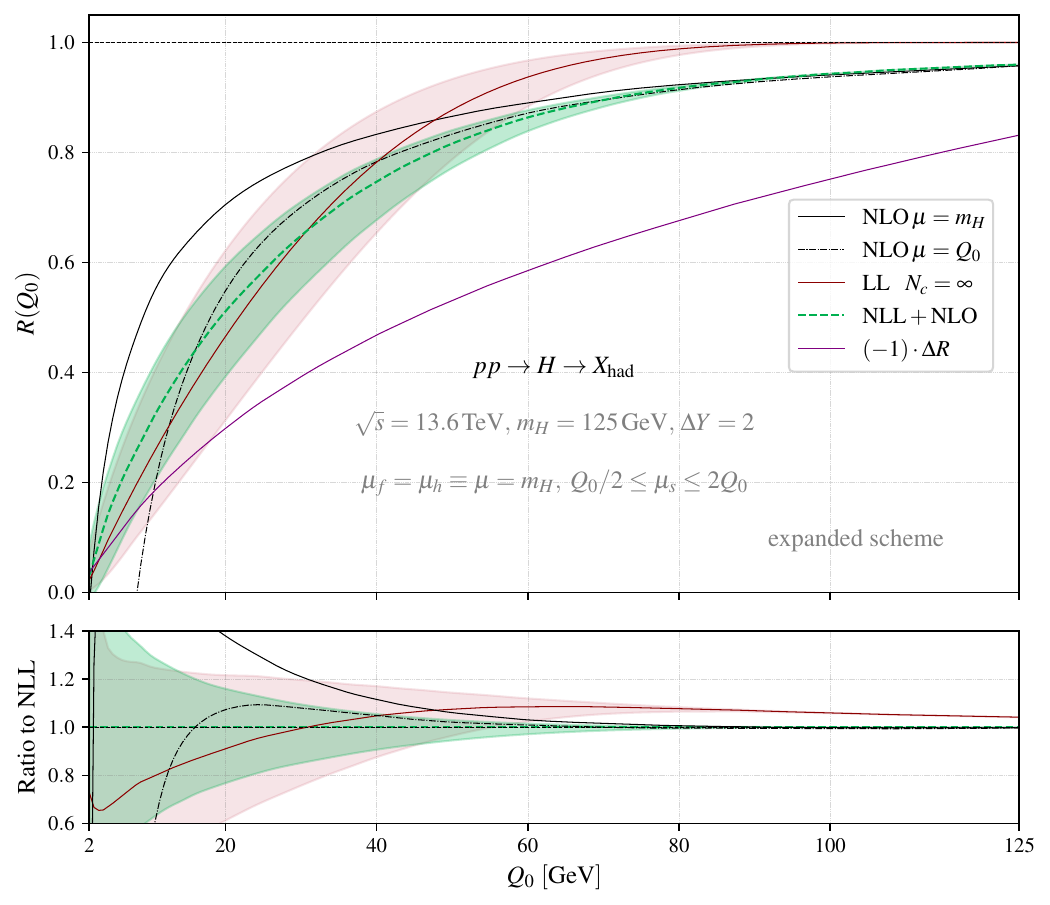}
        \caption{Expanded scheme}
    \end{subfigure}
    
    \vspace{0.5cm}
    \begin{subfigure}{0.48\textwidth}
        \includegraphics[width=\linewidth]{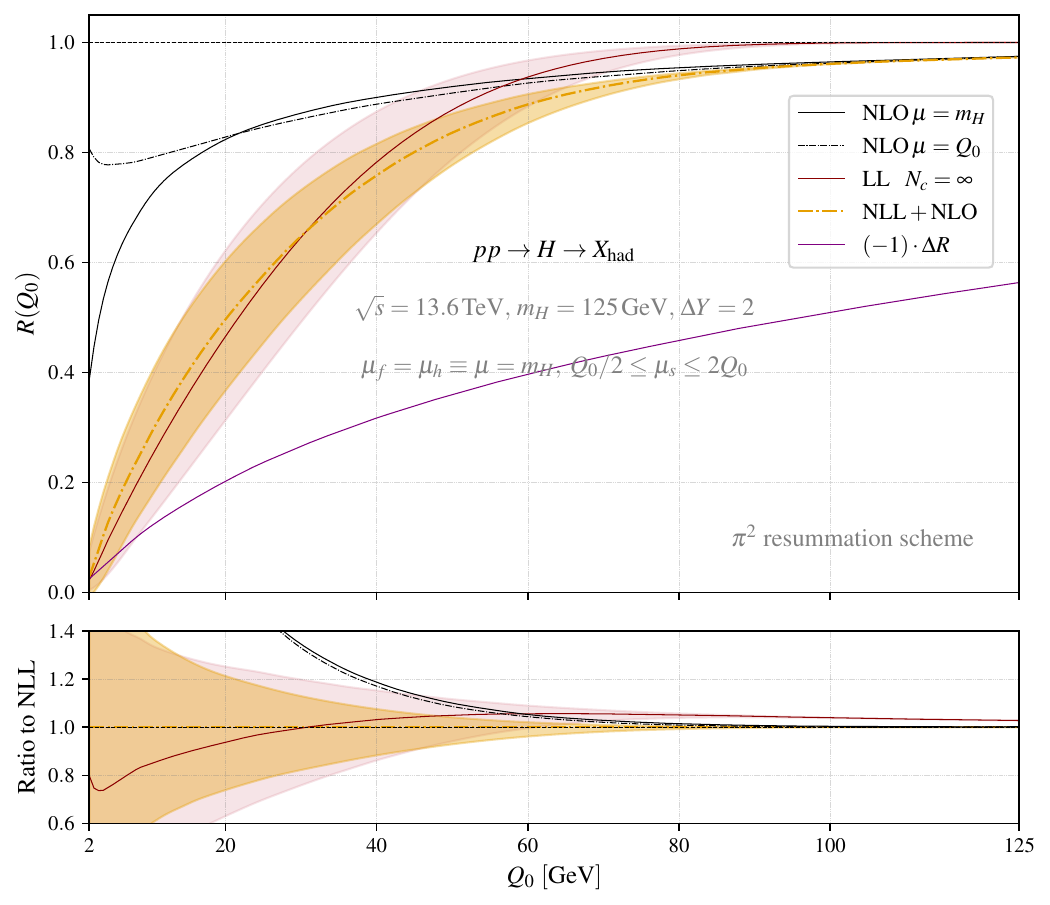}
        \caption{$\pi^2$ resummation scheme}
    \end{subfigure}
    \hfill
    \begin{subfigure}{0.48\textwidth}
        \includegraphics[width=\linewidth]{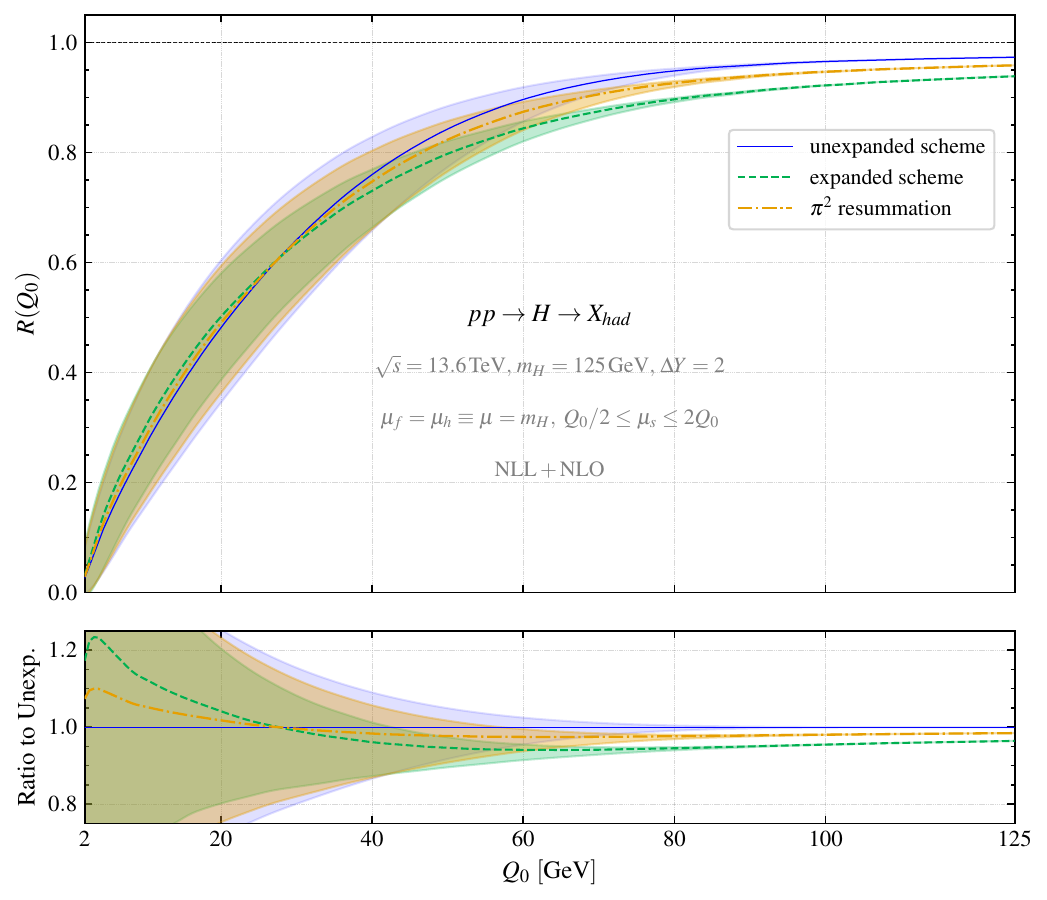}
        \caption{Comparison between the schemes.}
    \end{subfigure}
    \caption{The gap fraction $R(Q_0)$  at LL  and NLL+NLO under soft-scale variation $Q_0/2 \le \mu_s \le 2Q_0$. We show all three schemes as well as a comparison between the schemes at NLL+NLO.}
    \label{fig:gap_fraction_mH125}
\end{figure}

In Figure~\ref{fig:gap_fraction_mH125}, we show the resummed gap fraction for all three expansion schemes, together with a direct comparison plot among them.
The color coding follows the one used before: the LL prediction is shown in red, since it is identical for all schemes, while at NLL+NLO the unexpanded scheme is shown in blue, the expanded scheme in green and $\pi^2$ resummation in orange.
The fixed-order prediction is shown in black, solid for $\mu=m_H$ and dashed for $\mu=Q_0$.
The cancellation in the expanded scheme leads to significantly smaller uncertainties than for the unexpanded one.
For $\pi^2$ resummation, the uncertainties lie between those of the expanded and unexpanded schemes, as expected, since the $\pi^2$-enhanced terms only account for part of the inclusive corrections that do not cancel in the unexpanded ratio.
Despite the sizable differences in the scale uncertainties, the resummed predictions of the three schemes are in good agreement.

\begin{figure}
    \centering
    \begin{subfigure}{0.48\textwidth}
        \includegraphics[width=\linewidth]{Figures/GapFrac_mH125_musVary_xh1_unexpanded.pdf}
        \caption{Variation of the soft scale $\mu_s$}
    \end{subfigure}
    \hfill
    \begin{subfigure}{0.48\textwidth}
        \includegraphics[width=\linewidth]{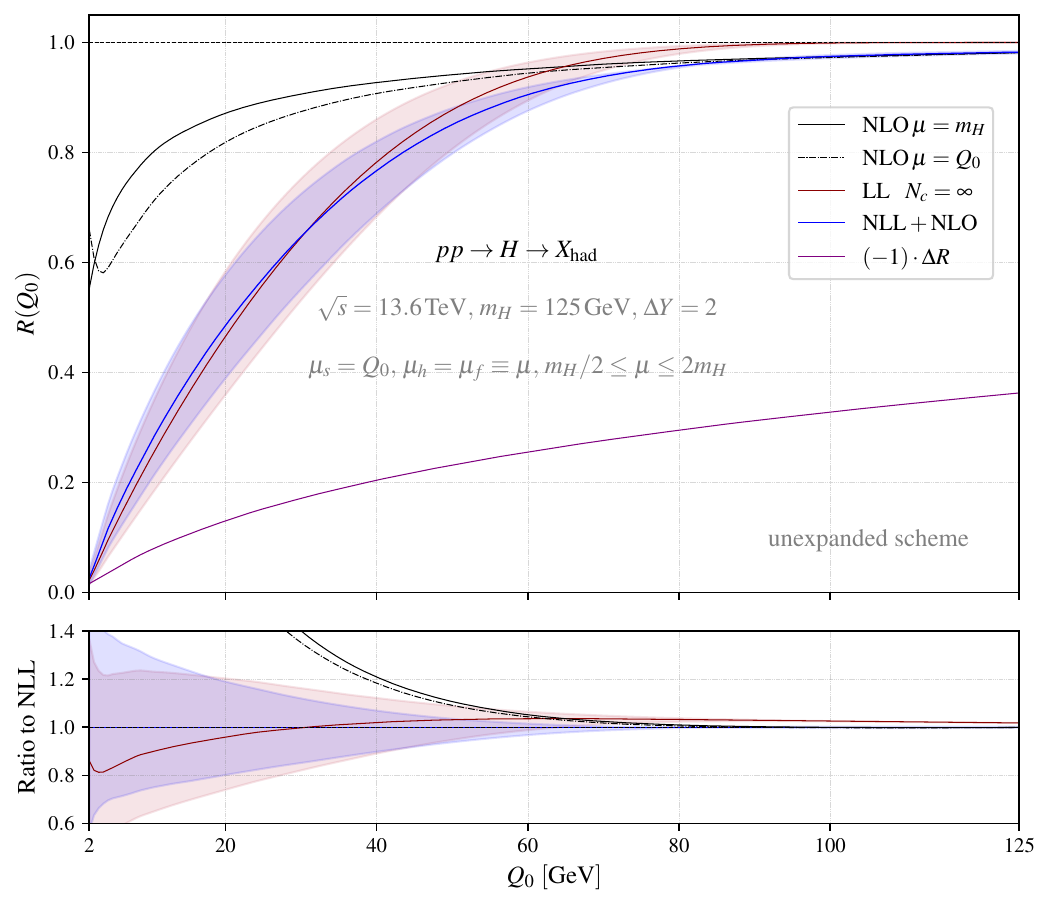}
        \caption{Variation of the hard scale $\mu_h$}
    \end{subfigure}
    \caption{The gap fraction $R(Q_0)$  at LL and NLL+NLO under (a) soft-scale variation and (b) hard scale variation.}
    \label{fig:gap_fraction_mus_vs_muh_vary}
\end{figure}

In Figure~\ref{fig:gap_fraction_mus_vs_muh_vary}, we assess the impact of soft and hard
scale variations, keeping the other scale fixed at its central value $\mu_h=m_H$ or
$\mu_s=Q_0$, respectively. The soft scale variation (left panel) clearly dominates over the hard scale variation shown in the right panel.
For the hard scale, the dependence partially cancels between the hard function
$\cH(\mu_h)$ and the soft function $\cS(t)$, which carries an implicit dependence on
$\mu_h$ through the shower evolution time $t(\mu_h,\mu_s)$.
The large soft scale dependence, in contrast, originates in the three-parton term
$\langle\cH_3^{(1)}\otimes\cS_3^{(0),\rm LL}(t)\rangle$, which has no one-loop soft
counterpart since $\cH_3$ already starts at relative order $\alpha_s$. The soft scale
enters this contribution only through the evolution time, without a compensating
matching correction.
We study this in more detail in
Appendix~\ref{app:scalevariations}, where we find that the dominant soft scale
uncertainty is generated by the hard matching contribution. Since the soft scale variations are larger than the hard scale variations, we will use the variation of the soft scale as our uncertainty estimate. All three expansion schemes show the same qualitative behavior for the soft and hard scale uncertainties, and we therefore refrain from showing hard scale variation plots for the other two schemes.

\begin{figure}[t]
    \centering
    \begin{subfigure}{0.48\textwidth}
        \includegraphics[width=\linewidth]{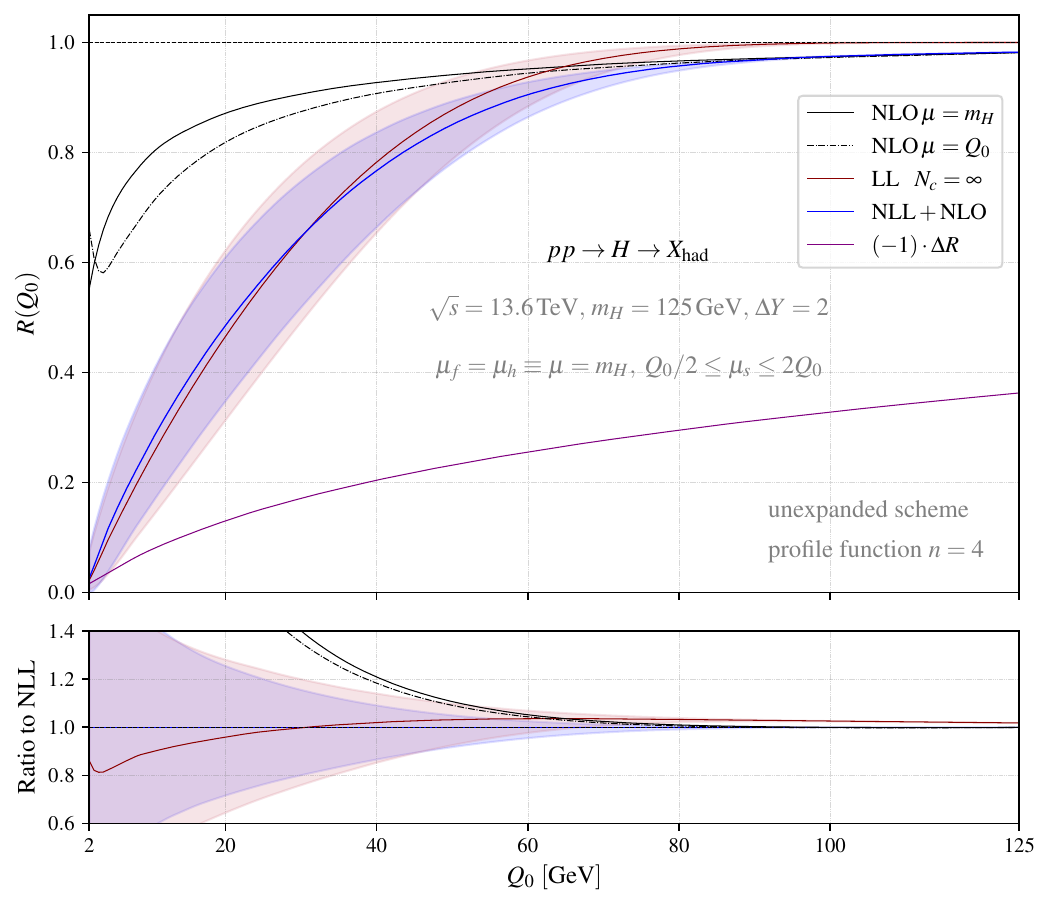}
        \caption{Profile function with $n=4$ (default)}
    \end{subfigure}
    \hfill
    \begin{subfigure}{0.48\textwidth}
        \includegraphics[width=\linewidth]{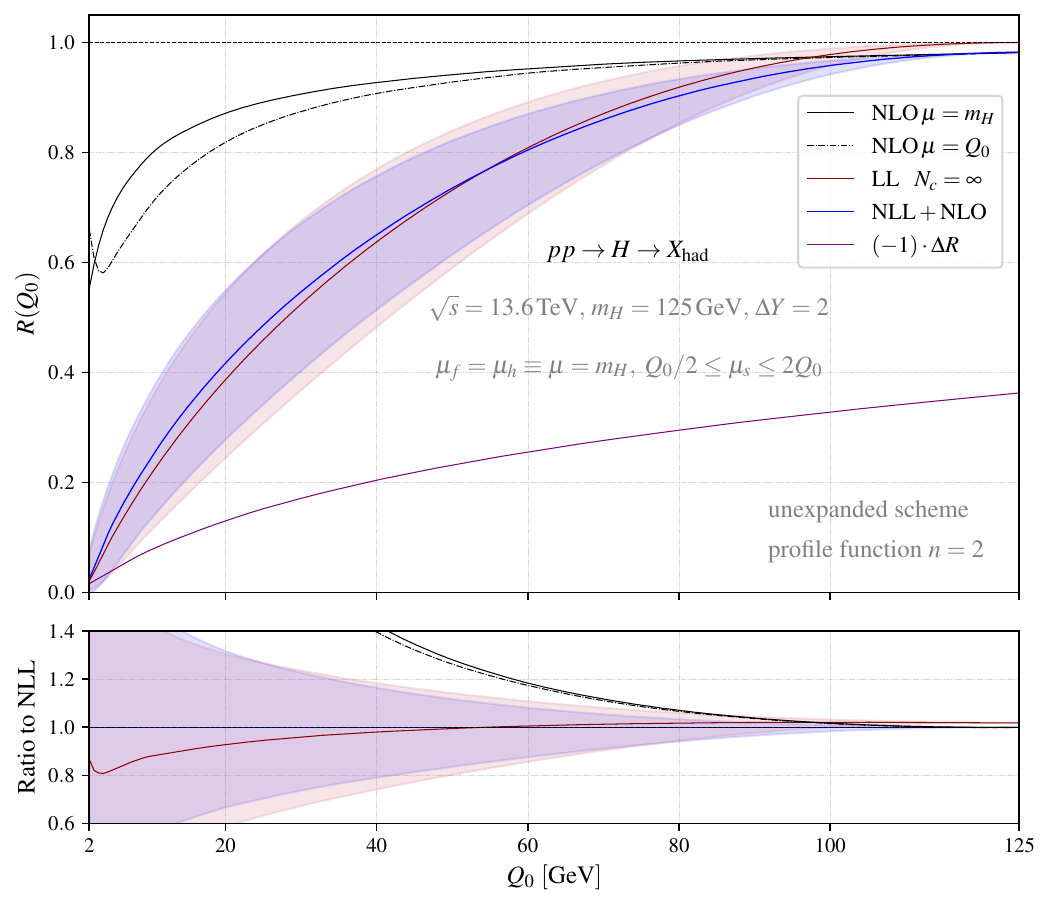}
        \caption{Profile function with $n=2$}
    \end{subfigure}
    \caption{The gap fraction $R(Q_0)$ at LL and NLL+NLO in the unexpanded scheme under soft-scale variation. Panel (a) shows the result using the default profile function with exponent $n=4$, while panel (b) illustrates the effect of modifying the profile function to use $n=2$.}
    \label{fig:gap_fraction_profile_comparison}
\end{figure}

Finally, we study the dependence on the profile function \eqref{eq:profile} introduced in Section~\ref{sec:matching2FO}.
Figure~\ref{fig:gap_fraction_profile_comparison} compares our LL and NLL+NLO predictions for $n=4$ (left) and $n=2$ (right).
For $n=4$, the resummation is switched off sharply, whereas $n=2$ leads to a more gradual transition to the fixed-order result.
At small $Q_0$, where the resummed prediction applies, the two choices agree within the scale uncertainties.
As $Q_0$ approaches $m_H$, the resummation is switched off for both choices, and they reduce to the fixed-order result by construction.
The dependence on $n$ therefore mainly affects the intermediate region, where we indeed find it to be strong.
While the choice of $n$ is to some extent arbitrary, a sharp transition is physically motivated: at $Q_0 \approx 80\,\text{GeV}$, the logarithm $\ln(m_H/Q_0)$ is no longer large and the fixed-order expansion is well behaved.
This supports our default choice $n=4$.

\section{Conclusion}\label{sec:conclusions}

In this paper we computed the Higgs cross section with a jet veto at central rapidities and resummed the leading and subleading non-global logarithms associated with the rapidity cut. The results were obtained using the {\sc Marzili} framework, which incorporates hard and soft matching corrections, as well as two-loop soft evolution. To capture the subleading logarithms, the Monte Carlo evolution also accounts for unordered soft emissions. Together with earlier results for Drell--Yan production \cite{Becher:2023vrh}, our results are the first resummations of subleading non-global logarithms in hadron-collider cross sections, and provide a benchmark for logarithmically accurate general-purpose parton showers. 

It is interesting to compare the Higgs production results to the Drell--Yan case, where we found the corrections to be moderate. In terms of soft functions and shower evolution, the two are closely related, and in the large-$N_c$ limit we can reconstruct the soft functions and their evolution for the Higgs case fully from the Drell--Yan results. Due to the larger color charge of the gluons, the correction terms associated with the running and the soft function matching are a factor of two larger than for Drell--Yan. The hard matching corrections for Higgs production are unrelated to the Drell--Yan case, and turn out to be by far the dominant correction. These large corrections are expected since they also arise in the inclusive Higgs cross section. In the presence of such large corrections, formally equivalent expansion schemes for the fraction of events that pass the veto at central rapidities lead to different results. In a fixed-order calculation, this leads to inconsistent event fractions, while the schemes are compatible after resummation.

In experimental measurements of the Higgs cross section with a jet veto $p_T^{\rm jet} < p_T^{\rm veto}$, the rapidity coverage is large enough that non-global effects are small for transverse momenta of $p_T^{\rm veto}\approx 30\,{\rm GeV}$. However, the situation is different for Higgs production in association with a single jet, or for Higgs production in vector-boson fusion. Both of these are genuinely non-global and to analyze these cross sections, the relevant hard matching corrections and the recently extracted two-loop clustering anomalous dimension \cite{Becher:2026zon} need to be implemented into {\sc Marzili}. We look forward to applying our resummation framework to these cases.

\subsection*{Acknowledgments}
NS is funded through the Royal Society, grant RP\textbackslash R\textbackslash 231001. TB has received funding from the Swiss National Science Foundation (SNSF) under grant 200021\_219377. RvK was supported by the Swiss High Energy Physics initiative for the FCC (CHEF) funded by SERI. XQ is supported by the Research Council of Finland, the Centre of Excellence in Quark Matter (project 364191).

\appendix
\section{Fixed-order validation and benchmarks}
\label{app:crosschecks}
\begin{table}
\centering
\begin{tabular}{lcccc}
\toprule
\raisebox{-2.1ex}[0pt][0pt]{$\left\{\frac{\mu_r}{m_H}, \; \frac{\mu_f}{m_H}\right\}$} & \multicolumn{2}{c}{This Work (pb)} & \multicolumn{2}{c}{\texttt{MCFM} (pb)} \\ 
\cmidrule(lr){2-3} \cmidrule(lr){4-5} 
                              & $\sigma_{\text{LO}}$ & $\sigma_{\text{NLO}}$ & $\sigma_{\text{LO}}$ & $\sigma_{\text{NLO}}$ \\ 
\midrule
$\{1, 1\}$                    & 13.9476              & 31.9681(3)               & $13.9476(4)$         & $31.9681(8)$ \\ 
$\{0.5, 1\}$                  & 17.2120              & 38.2723(3)               & $17.2120(5)$         & $38.2723(10)$ \\ 
$\{2, 1\}$                    & 11.5389              & 27.0989(2)               & $11.5389(3)$         & $27.0988(7)$ \\ 
$\{1, 0.5\}$                  & 13.1408              & 31.0524(3)              & $13.1408(4)$         & $31.0523(9)$ \\ 
$\{1, 2\}$                    & 14.5116              & 32.7039(2)               & $14.5115(4)$         & $32.7039(8)$ \\ 
\bottomrule
\end{tabular}
\caption{Comparison of the inclusive LO and NLO cross sections for the $gg \to H$ process at $\sqrt{s} = 13.6\,\text{TeV}$ between our results and MCFM. The absolute numerical integration errors for our leading-order implementation are below $10^{-4}$~pb and are therefore omitted in the table.}
\label{tab:scale_uncertainties}
\end{table}
We validate our expressions for the hard scattering kernels in~\eqref{eq:Cgg}--\eqref{eq:Cqq} by evaluating the hadronic cross sections numerically and comparing against predictions from \texttt{MCFM-10.3}~\cite{Campbell:2019dru}. 
In Table~\ref{tab:scale_uncertainties}, we give the inclusive NLO cross sections for Higgs boson production via gluon fusion
for different values of the renormalization and factorization scales, $\mu_r$ and $\mu_f$, where $\mu_r\equiv \mu_h$ in in the hard functions. As commonly done, we vary the scales by a factor of two around our central scale choice $\mu = m_H$. Choosing the same input parameters and PDFs as in Section~\ref{sec:pheno_predictions}, we find excellent agreement between our result and \texttt{MCFM}.

In addition to the inclusive cross sections presented in Table~\ref{tab:scale_uncertainties}, we tested the hard scattering kernels differentially by computing the transverse momentum distribution of the Higgs boson. 
A non-vanishing transverse momentum $p_T$ requires the Higgs boson to recoil against a real parton
such that the real-emission contributions from the hard scattering kernels contribute at leading order.
We compared our prediction for the transverse momentum spectrum against the LO $H+\text{jet}$ prediction from \texttt{MCFM} for $0.1 < p_T < 80\,\text{GeV}$ and find excellent agreement for all values of the transverse momentum with maximum deviations of 1\%.

The checks above validate the hard scattering kernels and their implementation. As a final consistency check, we verified that the factorized components assemble correctly into the vetoed cross section. Up to NLO, the latter can be evaluated in two independent ways: by direct numerical integration of the NLO phase space with the veto constraint $E_T < Q_0$, or by summing the individual factorized components (hard, soft, and matching). We compared the two for several scale choices and find agreement over the full range of $Q_0$, confirming that our factorized implementation reproduces the full NLO result.

\section{Additional scale-variations studies}
\label{app:scalevariations}

In this appendix we examine the scale dependence of our resummed predictions in more
detail. We first identify the origin of the large soft scale variation, then study the
impact of the central hard scale choice, and finally consider a heavy Higgs boson in
order to probe a larger hierarchy between $Q_0$ and $m_H$.

\begin{figure}
    \centering
    \includegraphics[width=0.7\linewidth]{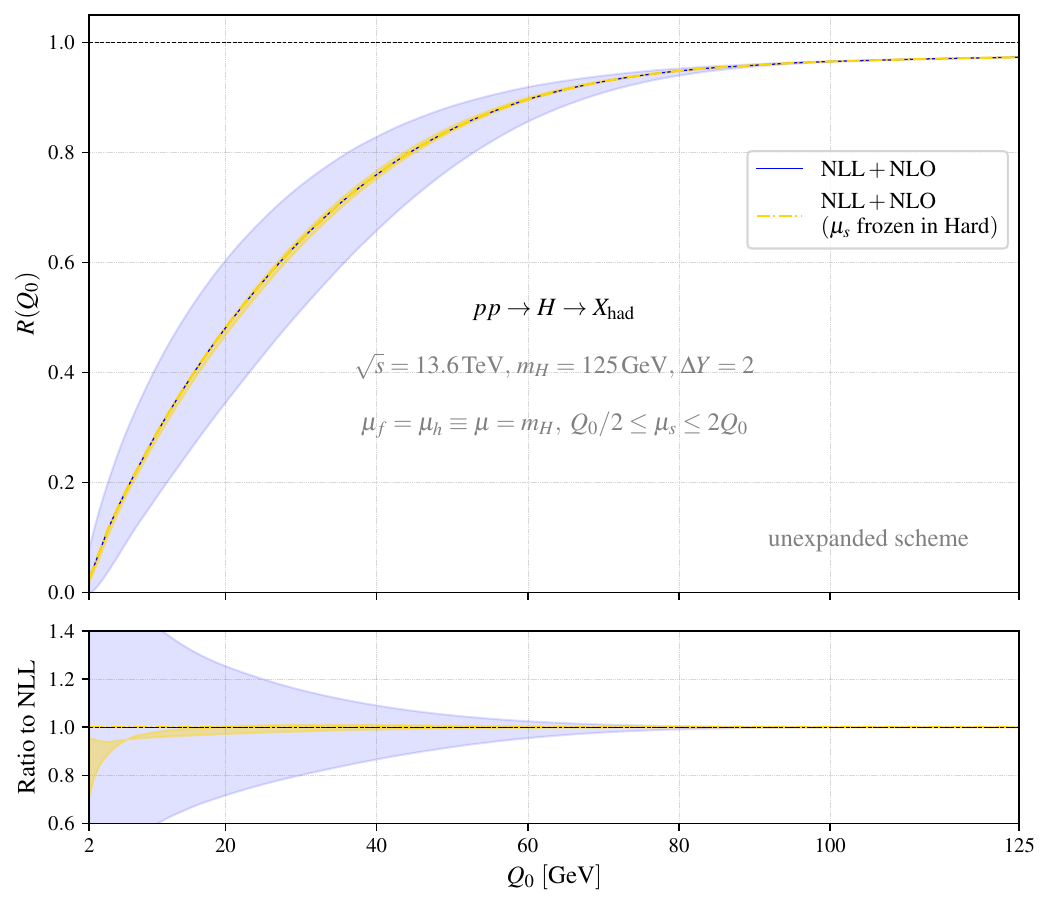}
    \caption{The gap fraction $R(Q_0)$ for $m_H = 125\,\text{GeV}$ evaluated in the unexpanded scheme ($\mu = m_H$) under soft scale variation $Q_0/2 \le \mu_s \le 2Q_0$. The default NLL+NLO prediction (blue) is compared against a modified setup where $\mu_s$ is frozen at $\mu_s=Q_0$ inside the NLO hard function (yellow, dash-dotted).}
    \label{fig:gap_fraction_band_check}
\end{figure}

Our resummed predictions carry sizable uncertainties even at NLL+NLO, and these are
dominated by the soft scale variation. To identify their source, we
compare in Figure~\ref{fig:gap_fraction_band_check} our default NLL+NLO prediction (blue)
with the same prediction evaluated with $\mu_s$ fixed to $Q_0$ inside the hard function
(yellow). Fixing the soft scale in the hard function yields a much narrower band. This is consistent with the expectation that the soft matching correction will cancel the soft scale dependence of the LL result up to higher-order terms. However, the soft-scale dependence of the NLO hard function will only get canceled at NNLL accuracy. Since the hard correction is large, we are left with a sizable soft scale uncertainty.

\begin{figure}
    \centering
    \begin{subfigure}{0.48\textwidth}
        \includegraphics[width=\linewidth]{Figures/GapFrac_mH125_musVary_xh1_unexpanded.pdf}
        \caption{unexpanded, $\mu_h = m_H$}
    \end{subfigure}
    \hfill
    \begin{subfigure}{0.48\textwidth}
        \includegraphics[width=\linewidth]{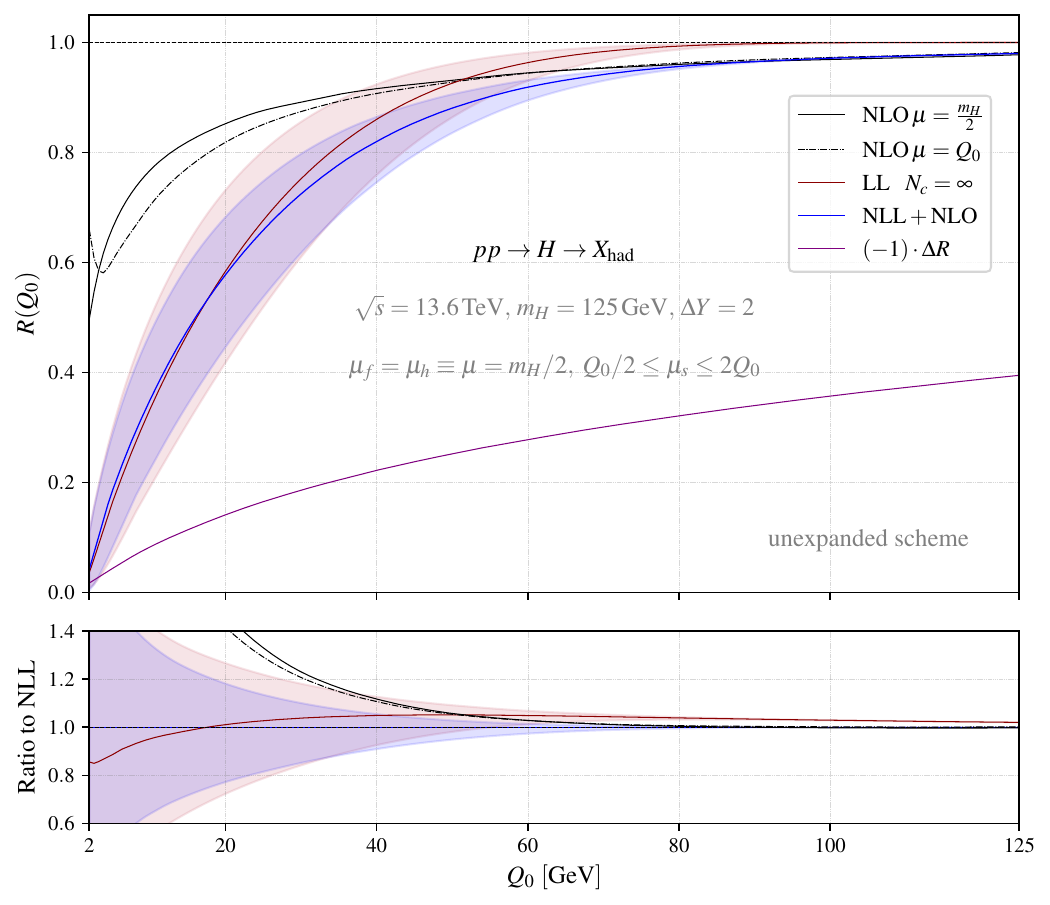}
        \caption{unexpanded, $\mu_h = m_H/2$}
    \end{subfigure}
    \caption{The NLL+NLO gap fraction $R(Q_0)$ for the central scale choices $\mu = m_H$ (left) and $\mu = m_H/2$ (right).
   }
    \label{fig:gap_fraction_musVary_mH125_mH05}
\end{figure}

We next turn to the choice of the central hard scale. In the context of gluon fusion, fixed-order predictions commonly adopt the low central
scale $\mu_r=\mu_f=m_H/2$. The motivation is largely empirical: the perturbative series
for the inclusive cross section converges considerably faster at this scale, with
successive scale-variation bands overlapping, whereas at $\mu=m_H$ the corrections at
each order remain sizable~\cite{Anastasiou:2015vya,Anastasiou:2016cez,LHCHiggsCrossSectionWorkingGroup:2016ypw}.
The underlying reason is that the inclusive cross section is dominated by soft and
collinear initial-state radiation off the gluons, whose characteristic scale lies well
below $m_H$.
A single-scale fixed-order
calculation must therefore compromise between the hard scale $\sim m_H$ and this much
lower radiation scale, and $\mu=m_H/2$ is the choice for which this compromise works
well numerically.
In our case no such compromise is necessary: the hard and soft
contributions are factorized, so the hard function is evaluated at its natural scale
$\mu_h=m_H$ while the soft scale is set independently.
For completeness we nevertheless
give the resummed gap fraction for the lower choice $\mu_h=m_H/2$.
Figure~\ref{fig:gap_fraction_musVary_mH125_mH05} compares our default $\mu_h=m_H$ (left
panel) with $\mu_h=m_H/2$ (right panel); as before, the uncertainty bands are obtained
by varying the soft scale. At NLL+NLO the lower choice yields a somewhat narrower band. We
note that this does not by itself indicate a more accurate prediction, as the band
reflects only the residual soft-scale dependence at this order. We show both predictions
in the unexpanded scheme only, since the other expansion schemes display the same
qualitative behavior.

\begin{figure}
    \centering
    \begin{subfigure}{0.48\textwidth}
        \includegraphics[width=\linewidth]{Figures/GapFrac_mH125_musVary_xh1_unexpanded.pdf}
      \caption{$m_H=125\,\rm GeV$}
    \end{subfigure}
    \hfill
    \begin{subfigure}{0.48\textwidth}
        \includegraphics[width=\linewidth]{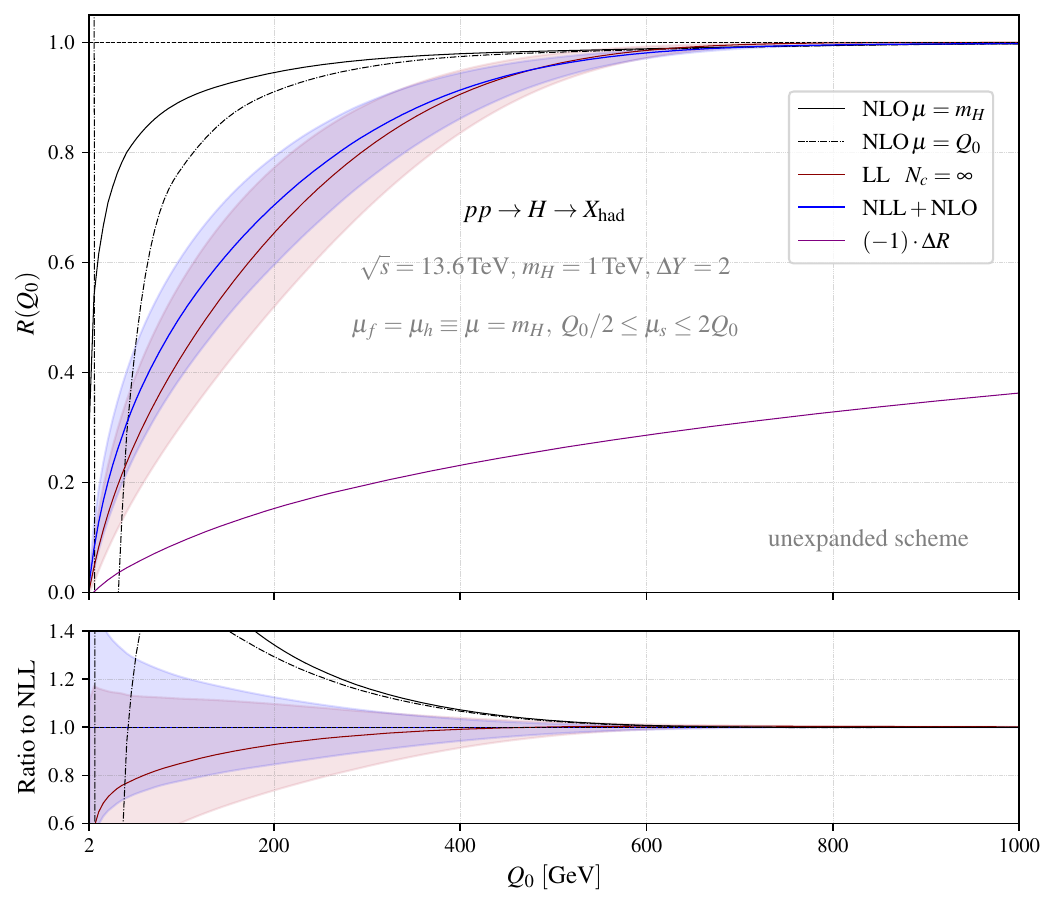}
      \caption{$m_H=1\,\rm TeV$}
    \end{subfigure}
    \caption{The NLL+NLO gap fraction $R(Q_0)$ for $m_H=125\,\rm GeV$ and $m_H=1\,\rm TeV$.}
    \label{fig:gap_fraction_musVary_mH125_mH1000}
\end{figure}

So far all results were obtained for the Standard Model Higgs boson. To study a larger
hierarchy $Q_0 \ll m_H$, we finally consider a fictitious heavy Higgs boson with $m_H = 1\,\rm TeV$.
In Figure~\ref{fig:gap_fraction_musVary_mH125_mH1000} we compare the gap fraction for the
two mass choices. Since the $Q_0$ range is scaled together with the Higgs mass, the two
predictions look very similar; the gap fraction is largely controlled by the ratio
$Q_0/m_H$. 
The remaining difference originates from the strong coupling, which is smaller
for $m_H = 1\,\rm TeV$ because it is evaluated at a higher scale.
The larger scale hierarchy has a more dramatic effect on the fixed-order prediction: for
$m_H = 1\,\rm TeV$, the result with $\mu = Q_0$ turns negative at small $Q_0$ before
rising again.

\section{NLL benchmark results}
\label{app:nll_benchmark}

To enable a future comparison of our results to other resummation frameworks, we now provide some benchmark numbers. To isolate the NLL correction, we
expand the gap fraction in $\as(m_H)$ while keeping
\begin{align}
 \lambda=\as(m_H)\ln\frac{Q_0}{m_H}
 \label{eq:benchmark_lambda}
\end{align}
fixed. We use the canonical scales $\mu_h=\mu_f=m_H$ and $\mu_s=Q_0$,
and hold the PDFs and hard kinematics fixed. In this limit, the LL terms
$\as^n\ln^n(Q_0/m_H)$ remain of order one, while the NLL terms
$\as^n\ln^{n-1}(Q_0/m_H)$ are linear in $\as(m_H)$. We thus retain the
resummed LL evolution and expand only its subleading corrections, in
contrast to the fixed-order expansion at fixed $Q_0/m_H$ in
\eqref{eq:matched}. Since $Q_0/m_H=\exp[\lambda/\as(m_H)]$ vanishes in
this limit for $\lambda<0$, power corrections do not contribute. For the benchmark, we therefore omit the nonsingular matching term and the profile scale.

With one-loop running, the evolution time \eqref{eq:evolution_time} and
the LL gap fraction are
\begin{align}
 t_{\rm LL}(\lambda)
 &=-\frac{1}{2\beta_0}\ln\left(1+\frac{\beta_0\lambda}{2\pi}\right),
 \qquad -\frac{2\pi}{\beta_0}<\lambda\leq0,
 \nonumber\\
 R_{\rm LL}(t)
 &=\mathcal S_{2,gg}^{(0),\rm LL}(t)
 =\left[\mathcal S_{2,q\bar q}^{(0),\rm LL}(t)\right]^2.
 \label{eq:benchmark_LL}
\end{align}
We define the relative NLL correction as
\begin{align}
 \frac{\Delta R_{\rm NLL}}{\as(m_H)R_{\rm LL}}
 =\lim_{\substack{\as(m_H)\to0\\\lambda\ {\rm fixed}}}
 \frac{R(\lambda,\as(m_H))/R_{\rm LL}(t_{\rm LL}(\lambda))-1}{\as(m_H)},
 \label{eq:benchmark_definition}
\end{align}
where $R$ is the leading-power NLL gap fraction. Thus
$\Delta R_{\rm NLL}$ is the term linear in $\as(m_H)$ in the expansion
of $R-R_{\rm LL}$ at fixed $\lambda$.

When the coupling is run at two loops, the evolution time itself
receives a subleading correction. It is convenient to combine it with
the part of the two-loop anomalous dimension proportional to the
one-loop kernel,
\begin{align}
 \bm\Gamma^{(2)}
 =\frac{\gamma_1^{\rm cusp}}{4}\bm\Gamma^{(1)}
   +\bm\Gamma^{(2)}_{\rm rem},
 \label{eq:benchmark_kernel_split}
\end{align}
where we use $\gamma_0^{\rm cusp}=4$ and
\begin{align}
 \beta_1=\frac{34}{3}C_A^2-\frac{20}{3}C_AT_Fn_f-4C_FT_Fn_f,
 \qquad
 \gamma_1^{\rm cusp}
 =4\left[\left(\frac{67}{9}-\frac{\pi^2}{3}\right)C_A
       -\frac{20}{9}T_Fn_f\right].
\end{align}
Following Section~6 of \cite{Becher:2026zon}, the cusp term and
the $-\beta_1\bm\Gamma^{(1)}/\beta_0$ term in \eqref{eq:NLL_insertion}
can be absorbed into the NLL evolution time
\begin{align}
 t_{\rm NLL}
 =\frac{1}{2\beta_0}\left[
 \ln\frac{\as(Q_0)}{\as(m_H)}
 +\frac{\as(Q_0)-\as(m_H)}{4\pi}
       \left(\frac{\gamma_1^{\rm cusp}}{4}-\frac{\beta_1}{\beta_0}\right)
 \right].
 \label{eq:benchmark_ttilde}
\end{align}
Evaluating $\as(Q_0)$ with two-loop running and expanding at fixed
$\lambda$, we obtain, with $t\equiv t_{\rm LL}(\lambda)$,
\begin{align}
 t_{\rm NLL}-t_{\rm LL}
 ={}&\frac{\as(m_H)}{4\pi}\frac{1}{2\beta_0}
 \left[
 2\beta_1t\,e^{2\beta_0t}
 +(e^{2\beta_0t}-1)
       \left(\frac{\gamma_1^{\rm cusp}}{4}-\frac{\beta_1}{\beta_0}\right)
 \right]
 +\cO(\as^2(m_H)).
 \label{eq:benchmark_time_shift}
\end{align}
The two terms in brackets arise from the two-loop running of the
coupling ratio and from the terms proportional to $\bm\Gamma^{(1)}$ in
the NLL insertion, respectively. The corresponding correction to the
gap fraction is
$\Delta R_{\rm run}=(t_{\rm NLL}-t_{\rm LL})R'_{\rm LL}(t)$, where the
prime denotes the derivative with respect to $t$. Since the Born state
contains two independent dipoles,
$R'_{\rm LL}=2\,\mathcal S_{2,q\bar q}^{(0),\rm LL}\,
\df\mathcal S_{2,q\bar q}^{(0),\rm LL}/\df t$, and at $t=0$ one finds
$R'_{\rm LL}(0)=-8N_c\Delta Y=-48$ for $N_c=3$ and $\Delta Y=2$.

The remaining contributions follow from \eqref{eq:NLL_master}
and~\eqref{eq:NLL_masterij}. After PDF convolution, we denote the hard
matching terms, summed over all partonic channels, by
$\Delta\sigma_{\rm hard}$, the one-loop soft matching term by
$\Delta\sigma_{\rm soft}$, and the insertion of
$\bm\Gamma^{(2)}_{\rm rem}$ into \eqref{eq:NLL_insertion} by
$\Delta\sigma_{\Gamma^{(2)}_{\rm rem}}$. In the latter, the
running-coupling weight remains $\as(t')=\as(m_H)e^{2\beta_0t'}$. All
three terms are evaluated at $t_{\rm LL}$, since shifting their time
argument would give an additional power of $\as(m_H)$. In the soft
matching term, the explicit coupling is evaluated at the low scale, so
that
\begin{align}
 \frac{\Delta\sigma_{\rm soft}}{\as(m_H)\sigma^{\rm LL}}
 =\frac{e^{2\beta_0t}}{4\pi}
       \frac{\mathcal S_{2,gg}^{(1),\rm LL}(t)}{R_{\rm LL}(t)},
 \label{eq:benchmark_soft}
\end{align}
with $\sigma^{\rm LL}=\sigma_{\rm incl}^{(0)}R_{\rm LL}$. The factor
$e^{2\beta_0t}=\as^{\rm LL}(Q_0)/\as(m_H)$ is of order one at fixed
$\lambda$ and therefore survives the expansion. The individual
contributions refer to the $\overline{\mathrm{LS}}$ scheme
\cite{Becher:2026zon}.

\begin{figure}[t]
 \centering
 \includegraphics[width=0.55\textwidth]{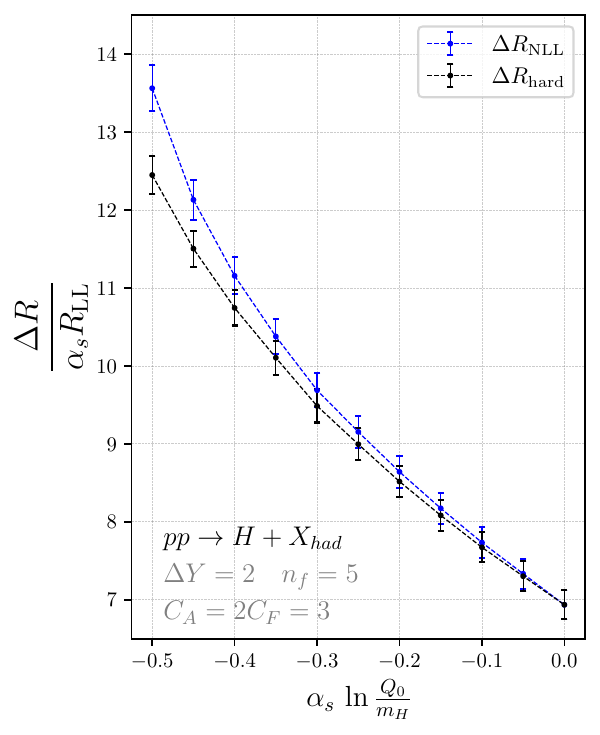}
 \caption{Relative NLL correction defined in
 \eqref{eq:benchmark_definition} for $\Delta Y=2$ and $n_f=5$.
 The blue curve shows \eqref{eq:benchmark_sum}, and the black curve
 shows the hard contribution including the inclusive subtraction.
 Both use $C_A=2C_F=3$, $m_H=125\,\mathrm{GeV}$ and the
 \texttt{NNPDF40\_nlo\_as\_01180} PDF set. The error bars denote
 numerical uncertainty estimates, including a $1\%$ relative allowance
 for residual cutoff dependence in the hard contribution.}
 \label{fig:nll_benchmark}
\end{figure}

Expanding the inclusive cross section as
$\sigma_{\rm incl}=\sigma_{\rm incl}^{(0)}+\as(m_H)\sigma_{\rm incl}^{(1)}
+\cO(\as^2(m_H))$, we obtain
\begin{align}
 \frac{\Delta R_{\rm NLL}}{\as(m_H)R_{\rm LL}}
 ={}&\frac{\Delta\sigma_{\rm hard}+\Delta\sigma_{\rm soft}
             +\Delta\sigma_{\Gamma^{(2)}_{\rm rem}}}
            {\as(m_H)\sigma^{\rm LL}}
   +\frac{\Delta R_{\rm run}}{\as(m_H)R_{\rm LL}}
   -\frac{\sigma_{\rm incl}^{(1)}}{\sigma_{\rm incl}^{(0)}}.
 \label{eq:benchmark_sum}
\end{align}
The last term in this equation is independent of $\lambda$ and arises because we analyze the correction to the gap fraction, rather than the correction to the cross section,
\begin{equation}
\frac{\Delta R_{\rm hard}}{\as R_{\rm LL}}
=\frac{\Delta\sigma_{\rm hard}}{\as\sigma^{\rm LL}}
-\frac{\sigma_{\rm incl}^{(1)}}{\sigma_{\rm incl}^{(0)}}\,.
\end{equation}
We will refer to this last contribution as the inclusive subtraction. It cancels common multiplicative
factors such as $|C_t|^2$ between the hard correction and the
inclusive cross section. The point $\lambda=0$ corresponds to the
continuation of the leading-power coefficient to $t=0$, rather than to
the matched gap fraction at $Q_0=m_H$. It need not vanish, since the
hard matching still retains the veto geometry.

For this benchmark we use leading-color hard coefficients, Born
normalization, inclusive subtraction and soft evolution, including
$\beta_1$, with $C_A=2C_F=3$ and $n_f=5$. The remaining inputs are the same as in Section~\ref{sec:pheno_predictions} and set $\mu_r=\mu_f=m_H$, together with $\Delta Y=2$. With
$\as(m_H)=0.112649$, we find
$\sigma_{\rm incl}^{(0)}=13.9476\,\mathrm{pb}$ and
$\as(m_H)\sigma_{\rm incl}^{(1)}=17.8681\,\mathrm{pb}$. The Born cross
section coincides with Table~\ref{tab:scale_uncertainties}, while the
leading-color NLO correction differs from the full-color value
$18.0205\,\mathrm{pb}$. The inclusive subtraction is
\begin{align}
 \frac{\sigma_{\rm incl}^{(1)}}{\sigma_{\rm incl}^{(0)}}\simeq11.3724.
 \label{eq:benchmark_inclusive_value}
\end{align}
With PDFs and hard kinematics fixed, this ratio is independent of the
reference coupling and is held fixed in the limit
\eqref{eq:benchmark_definition}.

\begin{table}[t]
 \centering
\begin{tabular}{lcc}
  \toprule
  \hspace{3mm}$\lambda$ & $\Delta R_{\rm NLL}/[\as R_{\rm LL}]$ & $\Delta R_{\rm hard}/[\as R_{\rm LL}]$ \\
  \midrule
  \hspace{1.8mm} $0.00$ &  $6.93(19)$ &  $6.94(19)$ \\
  $-0.05$ &  $7.33(19)$ &  $7.30(19)$ \\
  $-0.10$ &  $7.73(19)$ &  $7.67(19)$ \\
  $-0.15$ &  $8.17(20)$ &  $8.08(20)$ \\
  $-0.20$ &  $8.64(20)$ &  $8.52(20)$ \\
  $-0.25$ &  $9.15(21)$ &  $8.99(21)$ \\
  $-0.30$ &  $9.69(21)$ &  $9.49(21)$ \\
  $-0.35$ & $10.38(22)$ & $10.10(22)$ \\
  $-0.40$ & $11.15(24)$ & $10.74(22)$ \\
  $-0.45$ & $12.13(25)$ & $11.50(23)$ \\
  $-0.50$ & $13.56(29)$ & $12.45(24)$ \\
  \bottomrule
\end{tabular}
\caption{Numerical benchmark for the relative NLL correction in
 \eqref{eq:benchmark_definition}, with the parameters of
 Figure~\ref{fig:nll_benchmark}. All entries are dimensionless.
 Parentheses give absolute uncertainties in units of the last quoted
 digit, as in Table~\ref{tab:scale_uncertainties}. They combine the
 numerical uncertainty of the soft and evolution contributions with a
 $1\%$ relative allowance for residual cutoff dependence in the
 unsubtracted hard contribution, added linearly.}
 \label{tab:nll_benchmark}
\end{table}

The rapidity cutoffs are $\eta_{\rm cut}=8$ for the soft evolution and
$\eta_{\rm cut}=5$ for the hard integration. At finite cutoff, the hard
matching contribution retains a residual dependence on the rapidity
regulator, which we estimate by assigning a $1\%$ relative uncertainty
to $\Delta\sigma_{\rm hard}$ before the inclusive subtraction. This
prescription is motivated by the cutoff studies in Section~6 of
\cite{Becher:2026zon}. After the inclusive subtraction \eqref{eq:benchmark_inclusive_value}, the relative hard-correction uncertainty is around 2\% and is added linearly to the numerical uncertainty of the soft and evolution contributions. If necessary, the hard function uncertainty can be reduced at the expense of additional runtime.

The results are shown in Figure~\ref{fig:nll_benchmark} and listed in
Table~\ref{tab:nll_benchmark}. Even after the inclusive subtraction \eqref{eq:benchmark_inclusive_value}, the hard correction
dominates over the entire range of $\lambda$. The negative soft and
running contributions partially cancel the positive contribution of
$\bm\Gamma^{(2)}_{\rm rem}$, so that their sum remains below $10\%$ of
the total correction.

\bibliographystyle{JHEP}
\bibliography{bibliography}

\end{document}